\documentclass[]{article}

\usepackage{astron}

\usepackage{supertabular}
\usepackage{longtable}
\usepackage{graphicx}

\def\degr{\hbox{$^\circ$}}

\newenvironment{changemargin}[2]{%
 \begin{list}{}{%
  \setlength{\topsep}{0pt}%
  \setlength{\leftmargin}{#1}%
  \setlength{\rightmargin}{#2}%
  \setlength{\listparindent}{\parindent}%
  \setlength{\itemindent}{\parindent}%
  \setlength{\parsep}{\parskip}%
 }%
\item[]}{\end{list}}

\begin{document}
%\twocolumn
%\begin{article}
%\begin{opening}
%\null\vskip2cm
%\begin{center}

\title{The Oort spike modification due to non-gravitational effects}
\author{Ma{\l}gorzata Kr{\'o}likowska\footnote{email: mkr@cbk.waw.pl}\\
Space Research Centre of the Polish Academy of Sciences\\
Bartycka 18A, 00-716, Warsaw, Poland}

\maketitle

\begin{center}
%{\small submitted to AA}

\end{center}

\abstract{This is the third of a series of papers on investigating
the non-gravitational effects in the motion of long-period comets.
The influence of the non-gravitational effects on the original and
future orbital elements, and in particular on reciprocals of
semimajor axes is analyzed.The new orbit determinations were
performed in a homogenous way using the  basic collection of 50
nearly-parabolic comets of first class orbits. The data consist of
23 hypothetically hyperbolic comets (sample~I), 8~Oort spike comets
(sample~II), and 19~'old' comets (sample~III). The original and
future reciprocals of semimajor axes for each comet were calculated
for strictly gravitational case and the non-gravitational case in a
fully consistent way. Clear shift of the original reciprocals of
semimajor axes toward the more elliptical orbits is demonstrated.
The average values of the original and future energy changes due to
the non-gravitational acceleration are derived for each of three
samples independently. Omission of the NG~effects significantly
affects the position and width of the Oort spike. The shift of the
Oort spike position from the standard $1/a_{\rm ori}$ value of
2--4~$\cdot 10^{-5}$\,AU$^{-1}$ to the value greater than $5.5\cdot
10^{-5}$\,AU$^{-1}$ is discussed.}

% \keywords{Solar system -- comets: long-period; non-gravitational effects}

%\setcounter{table}{2}

\section{Introduction}

More than half century ago Oort \cite*{Oort} postulated that at
distances of tens of thousands AU existed roughly spherical cloud of
comets. He has built his famous idea on just 19 well determined
cometary orbits. This classical vision of Oort cloud remains still
valid on the basis of almost four hundred
%high-quality
orbits of nearly-parabolic comets discovered up to now (Marsden and
Williams Catalogue of Cometary Orbits, \cite*{MWC}, hereafter
MW~Catalogue)\footnote{Almost all orbits of nearly parabolic comets
are derived as purely gravitational}. Almost 30\% of long-period
(LP) comets have $E_{\rm ori}\equiv 1/a_{\rm ori} <
10^{-4}$\,AU$^{-1}$ where about one fourth of them
%(38~objects, see Tab.~\ref{tab:new})
have negative $E_{\rm ori}$. However, recent determinations of the
typical $E_{\rm ori}$ gave significantly smaller values than
original indicated by Oort. Oort estimated that most of nearly
parabolic comets have semimajor axes of about 25--75 thousands AU.

%\footnote{The sign convention for binding energy, E, is the opposite
%of that used for orbital energy per unit mass}
%$-{\rm G}\cdot {\rm M}_{\odot }\cdot (2{\rm a})^{-1}$

The median semimajor axis of the 109~comets with the first quality
orbits and $0 < E_{\rm ori} < 10^{-4}$\,AU$^{-1}$ selected from the
MW~Catalogue is 27\,800\,AU where all orbits are strictly
gravitational. If all available non-gravitational (NG) orbits of $0
< E_{\rm ori} < 10^{-4}$\,AU$^{-1}$ are used the sample increases to
120~comets and the median semimajor axis declines to 26\,000\,AU.
This result strongly suggests that taking the NG~effects into
account may reduce the Oort Cloud distance from the Sun
significantly below 50\,000\,AU.

In the above estimates the comets with a negative value of the
binding energy were excluded. These objects could be interstellar
intruders passing through the inner solar system. However, the
modern estimates of the expected rate for extrasolar comet detection
give at most a few on average every 450 yrs \cite{SeR93,Hug91}. Thus
the majority of 'hiperbolic' comets have to have local origin. This
discrepancy between observations and theoretical expectations
disappears when the NG~effects are included for orbit determination
(Marsden et al. \cite*{MSY,MSE}, Kr{\'o}likowska
\cite*{Kro1}(Paper~I) \cite*{Kro2}(Paper~II)). It is due to a fact,
that NG~forces make 'hiperbolic' and elliptical osculating orbits on
the average less eccentric than in the case of purely gravitational
motion. Consequently, omission of the NG~effects may significantly
affect the position and width of the Oort spike. It means also that
the apparent inner edge of the Oort cloud should be shifted towards
the more positive $1/a_{\rm ori}$. In the present paper I try to
give quantitative estimates for the magnitude of these shifts.

In Paper~I the sample of 33 comets with negative value of
reciprocals of original semimajor axis were analyzed and for 16 of
them the NG-effects could be determined. It was shown that for 14 of
these 16 comets the incoming barycentric orbits changed from
hyperbolic to ecliptic when the NG~effects were included. In
sections~2~and~3 of the present paper the analysis of the NG~effects
for the almost complete sample of comets that have the catalogue
value\footnote{By catalogue value I mean the value given in
MW~Catalogue of Cometary Orbits, 2005} of $1/a_{\rm ori} <
10^{-5}$AU$^{-1}$ is performed. For those comets the term
'dynamically new' (in short 'new') is applied, while comets with
$10^{-5}$AU$^{-1}< 1/a_{\rm ori} < 10^{-4}$AU$^{-1}$ are defined as
the 'Oort spike' comets. This differentiation between 'new' and
'Oort spike' comets, however, does not imply that the Oort spike
comets cannot be dynamically new objects. In fact, simulations by
Dybczynski \cite*{Dyb05} have shown that the value of $4\cdot
10^{-5}$AU$^{-1}$ suggested previously by Marsden and Sekanina
\cite*{MS} is the most appropriate for definition of dynamically new
comets.  In Section~4 two other samples of LP comets with NG~effects
have been constructed. The first sample contains comets with the
catalogue orbit inside the Oort spike (8 objects), the second --
comets with catalogue $1/a_{\rm ori}
> 10^{-4}$AU$^{-1}$ (19 objects). It is important to stress that the
NG~effects were derived simultaneously with osculating orbital
elements from the positional data taken near perihelion passage and
were then consistently applied to determine the original/future
orbits and original/future reciprocals of semimajor axis. The
variations of the original and future energy due to the NG~effects
are widely discussed in Section~6, where the average values of
differences between the original orbital energies derived from
NG~orbit and strictly gravitational orbit, $\Delta E_{\rm ori} =
1/a_{\rm ori, NG} - 1/a_{\rm ori, GR}$, were determined separately
for all three basic samples. These average values of $\Delta E_{\rm
ori}$ allow to correct the observed Oort spike for the NG-effects
(section~7).

Modern simulations of the Oort Cloud evolution predict that the
observed sample of new comets that reach the inner solar system is
biased to objects with the semimajor axes greater than
20\,000--30\,000\,AU because of the 'Jupiter barrier'. New comets
must have decreased their perihelion distances from more than 10\,AU
to less than 3\,AU in one orbital period. Otherwise, they had a
chance to encounter Jupiter and/or Saturn during their earlier
evolution, which would remove them from the Oort spike. Heisler
\cite*{He90} has derived that the binding energies of Oort spike
comets are peaked at $E_{\rm ori} = 3.5\cdot 10^{-5}$\,AU$^{-1}$ ($a
\sim 29\,000$\,AU), assuming a local density of
0.185\,M$_{\odot}$/pc$^3$. Taking currently accepted value of
0.1\,M$_{\odot}$/pc$^3$ for local density, Levison et al.
\cite*{Lev01} argued that the maximum should be rather at $\sim
$~34\,000\,AU, whereas the inner edge of the outer Oort cloud -- at
$\sim $28\,000\,AU. Thus, one should notice some discrepancy between
the theoretical predictions and apparent inner edge at 20\,000\,AU.
However, both the theoretical calculations and orbit determinations
do not include the NG-effects. Hence it is open question if this
discrepancy is real \cite{Ric04}.

\begin{table*}
\begin{center}
\caption{{\small The samples of the investigated LP comets with
first quality orbits}}
% The asterisk$^*$ denotes that the most 'hyperbolic' comet C/1853~R1~Bruhns with
%poorly determined catalogue orbit was removed from the sample~A and AA
\label{tab:3sa} \vspace{0.10cm}
%{\setlength{\tabcolsep}{1.0mm}
{\small
\begin{tabular}{cccc} \hline \hline
%\begin{tabular}{ccc}
%~\hspace*{5.5cm}& {\hspace*{4.0cm}}           & {\hspace*{4.0cm}}      \\
Comets             & designation & criterion                                  & Number     \\
                   &             &                                            & of comets \\
 \hline
% &&&  \\
\multicolumn{4}{c}{\bf Basic samples (contain comets with the detectable NG effects)} \\
'new'              & I           & $1/a_{\rm ori} < 10^{-5}$\,AU$^{-1}$                      & 23         \\
Oort spike         & II          & $10^{-5}$\,AU$^{-1} < 1/a_{\rm ori} < 10^{-4}$\,AU$^{-1}$ &  8         \\
'old'              & III         & $1/a_{\rm ori} > 10^{-4}$\,AU$^{-1}$                      & 19         \\
 \hline
% &&&  \\
\multicolumn{4}{c}{\bf Global samples constructed directly from the MW Catalogue;} \\
\multicolumn{4}{c}{ contain LP comets with aphelions $Q > 250$\,AU }              \\
'new'+Oort spike   & A           & $1/a_{\rm ori} < 10^{-4}$\,AU$^{-1}$           & 131        \\
'old'              & B           & $1/a_{\rm ori} > 10^{-4}$\,AU$^{-1}$           & 131        \\
%all                & C           &                                                & 376             \\
 \hline
% &&&  \\
\multicolumn{4}{c}{\bf Corrected  global samples} \\
\multicolumn{4}{c}{ contain LP comets with aphelions $Q > 250$\,AU }              \\
'new'+Oort spike   & AA          & $1/a_{\rm ori} < 10^{-4}$\,AU$^{-1}$           & 144$-$7  \\
'old'              & BB          & $1/a_{\rm ori} > 10^{-4}$\,AU$^{-1}$           & 142$+$7  \\
%all                & CC          &                                                & 398      \\
 \hline
\end{tabular}}

%}}
\end{center}
\end{table*}

\begin{figure}
%\vspace{0.3cm}
\begin{center}
\includegraphics[width=9.0cm]{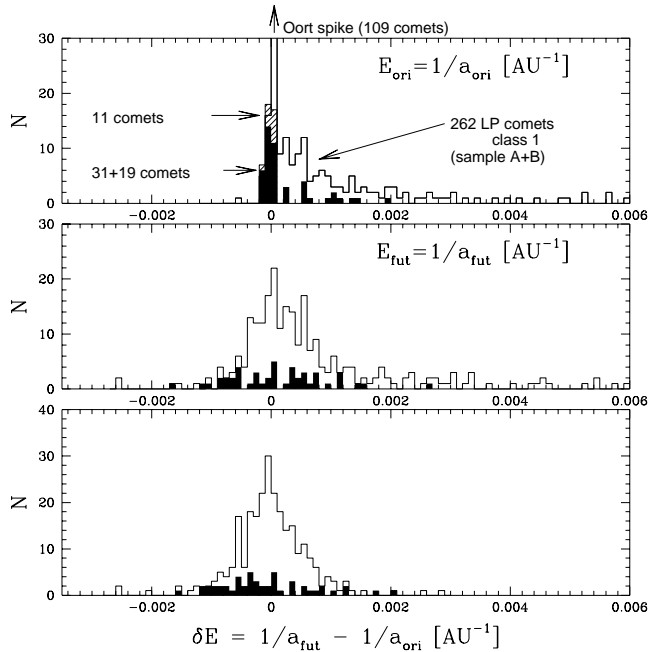}
\end{center}
\caption{{\small Distributions of cometary energies, $E_{\rm ori}
\equiv 1/a_{\rm ori}$ (top panel), and $E_{\rm fut} \equiv 1/a_{\rm
fut}$ (middle panel), $\delta E\equiv 1/a_{fut} - 1/a_{\rm ori}$
(bottom panel). The filled histograms show the respective energy
distributions derived for strictly gravitational orbits for all of
the investigated comets (50 objects from all three samples:
I+II+III; Table~\ref{tab:3sa}). The shaded histogram drawn in the
top panel represents the 11 comets with $E_{\rm ori} <
10^{-5}$\,AU$^{-1}$, where the NG-effects were indeterminable (part
(i) of Table~\ref{tab:new}; sample AA). The solid line histograms
represent global distributions of all comets with first quality
orbits and aphelions $Q>250$\,AU taken from MW~Catalogue
(sample~A+B, Table~\ref{tab:3sa})}}\label{fig:376}
\end{figure}

%%%\end{document}

\section{Sample of hypothetic 'hyperbolic' comets}

The MW~Catalogue contains 404 'single-apparitions' nearly-parabolic
comets of the first and second quality orbits in the MW~Catalogue
with aphelion greater than 250\,AU. Among them 29 comets have the NG
orbits, and these objects are analyzed separately.

To calculate the original and future reciprocals of the semimajor
axes, each of 375 ($404-29$) comets were followed from their
position at a given epoch backwards (incoming orbit before entry
into the planetary system) and forwards (outgoing orbit after
planetary perturbations) until the comet reached a distance of 250
AU from the Sun. The equations of motion have been integrated in
barycentric coordinates using the recurrent power series method
(Sitarski \cite*{Si89,Si02}) taking into account the perturbations
by all nine planets. All the numerical calculations presented here
are based on the Warsaw ephemeris DE405/WAW, i.e. the numerical
Solar System ephemeris consistent to high accuracy with the JPL
ephemeris DE405 \cite{Si02}.

The sample of 44 comets with original value of energy $E_0 \equiv
1/a_{\rm ori} < 10^{-5}$\,AU$^{-1}$ was extracted from the 375 sets
of the orbital elements taken from the MW~Catalogue~(2005). The
derived values of $1/a_{\rm ori}$ were generally very similar to
those given in the MW~Catalogue. However, three comets in part~(i)
of Table~\ref{tab:new} have catalogue values of $1/a_{\rm ori}$
greater than $10^{-5}$\,AU$^{-1}$ (C/1978~R3, C/1987~W3, C/2002~A3).

%(column 6 of Table\ref{tab:new}a
%(column 5 of Table\ref{tab:new}a)

Among these 44~objects, one is the secondary component of the
splitting event (C/1996 J1A). It is obvious that original orbits, in
particular the value of $1/a_{\rm ori}$, would be unreliable for
this fragment.
%Only the leading component of C/1996~J1 splitting satisfied the
%selection criteria and is included (see Table~\ref{tab:new}A).
The leading component of this splitting is included (part (i) of
Tables~\ref{tab:new}). The observations for the three comets
(C/1849~G2, C/1906~B1, C/1959~O1) were not available. Four recently
discovered comets (C/2003~T3, C/2005~B1, C/2005~G1 and C/2005~K1)
are still potentially observable and were not taken into account.
The resulting sample of 36 comets was supplemented by {eleven}
comets with NG-orbits in the MW~Catalogue for which $1/a_{\rm ori} <
10^{-5}$\,AU$^{-1}$ were obtained for strictly gravitational orbits
determined directly from observations.
%({\bf part (ii)} of Tables~\ref{tab:obs} and \ref{tab:new}).

Thus, the new orbit determination was performed for 47 objects of
the sample of 'dynamically new' comets, where 34 have the first
class orbits. Seven comets in this sample (C/1959~Y1, C/1986~P1A,
C/1990~K1 C/1993~A1, C/1995~Y1, C/1996~E1, C/1998~P1) were
previously analyzed in Paper~II and their osculating orbits for
strictly gravitational model and standard NG-model presented there
are incorporated into the present data. Of the remaining 40 objects,
27 comets have been investigated in Paper~I. In the present paper
for eight of them the observational material has been significantly
extended. I additionally collected all observations available in the
literature for six comets that were discovered at the end of XIXth
century. For the eight comets observed during the period 1900-1950
the analysis was based using the data collected in Warsaw in
cooperation with the Slovakian group at the Astronomical Institute
from Bratislava and Tatranska Lomnica. Generally, the positional
data used for the orbit determination in the present work are more
complete than the data described in the MW Catalogue. The
observational arcs and number of observations
%and {\it rms}
taken for the strictly gravitational and NG orbital fitting to the
positional data  are collected in Table~\ref{tab:new} (parts (i) and
(ii)) for all the comets with the first class orbits (34~objects) as
well as for six comets with the second class orbits and with
determinable NG~effects (6~objects). Seven comets with the second
class orbits have indeterminable NG~effects and these objects are
not presented in Table\ref{tab:new}.

The original and future reciprocals of semimajor axis obtained from
the new orbit determinations of these 40~'new' comets are given in
parts~(i) and~(ii) of Table~\ref{tab:new} (columns 7--10) The basic
sample~I was constructed from the 'new' comets having first quality
orbits and the detectable NG~effects (column~6 of
Table~\ref{tab:new}).

%One can see that three comets have $1/a_{\rm ori}$ more
%negative than $-0.000200$\,AU$^{-1}$. Among them is C/1955~O1~Honda where the
%splitting was observed.

\section{Comparison between new and previous orbit determinations}

New determinations of osculating orbits and original and future
reciprocals of semimajor axis (Table~\ref{tab:new}, columns~6--~9)
differ in a few cases from those given in Paper I (Tables~1 and 2
therein) due to three reasons.

First, the observational material for several comets is more
complete than in Paper~I. Second, I have applied the different
selection procedure that was successfully verified using the
long-period comets with NG-effects (Paper II), where the
pre-perihelion and post-perihelion observations were selected
separately.

This new approach and larger observational data have allowed to
determine the NG-effects for six comets listed in Paper~I among
objects with the indeterminable NG-effects (C/1899 E1, C/1952~W1,
C/1892 Q1, C/1996 J1B, C/1997 BA6, C/1997 J2). Moreover, residuals
for another six comets investigated also in Paper~I are greater than
those given there. In all cases the new selection procedure removes
smaller sets of positional observations in rectascension or/and in
declination than the method applied in Paper~I.

Finally, the present numerical calculations are based on the Solar
System ephemeris DE405/WAW, whereas the results given in Paper I
were based on the older one. For more detailed discussion see
Paper~II.

%Comparing strictly gravitational and the NG solutions for
%'hyperbolic' comets given in Table~1 of Paper~I with those given
%here in Table~\ref{tab:new} one can see that values of $1/{\rm
%a}_{\rm ori}$ are very sensitive to the selection procedure.
%However, the difference $1/a_{\rm ori,NG}-1/a_{\rm
%ori,GR}$ for any individual objects are always very similar. Thus,
%the analysis based on differences in original or future reciprocals
%of semimajor axes is less sensitive to the selection procedure and
%numerical method used in an investigation of LP~comets.

\begin{figure}
%\vspace{0.3cm}
\begin{center}
\includegraphics[width=9.0cm]{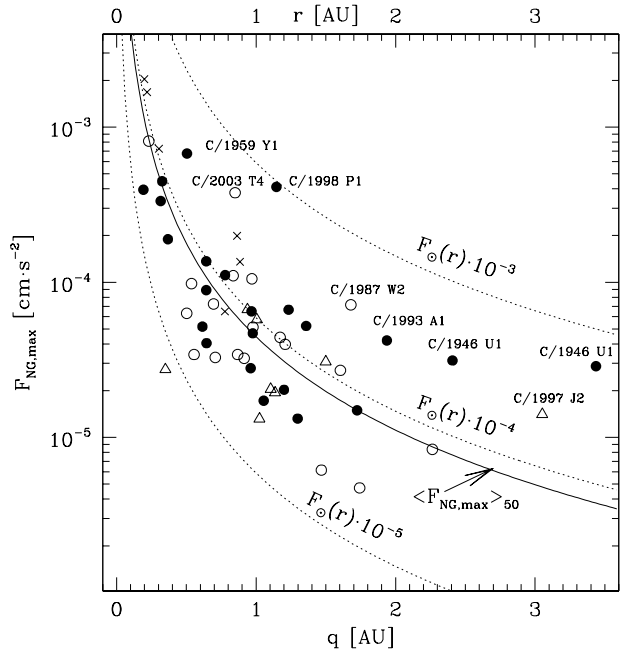}
\end{center}
\caption{{\small Maximum values of the NG~forces $F_{\rm{NG,max}}=
A_{\rm 1}\cdot {\rm g}(r=q)$ vs. perihelion distance, $q$, derived
for the investigated LP~comets. Full circles, open triangles, and
open circles  show maximum values of the NG-forces for the sample I,
II, and III, respectively. The crosses represent six comets with
second quality class orbits given in part(i) of Table~\ref{tab:new}.
The dotted curves represent the $10^{-3}, 10^{-4}, 10^{-5}$ of the
solar gravitational acceleration $F_{\odot}({\rm r})$, respectively,
while the solid curve shows the mean value of $F_{\rm{NG,max}}$
derived from the sample I+II+III}} \label{fig:NGforce}
\end{figure}

\section{Two other samples of comets with detectable NG effects}

I have constructed two another samples of comets with the detectable
NG-effects and the first quality orbits: Oort spike comets
($10^{-5}$\,AU$^{-1} < E_{\rm ori} < 10^{-4}$\,AU$^{-1}$; 8~objects)
and 'old' comets ($E_{\rm ori} > 10^{-4}$\,AU$^{-1}$; 19 comets).
The observational material of both samples and the original and
future reciprocals of semimajor axis are given in
Table~\ref{tab:new}. In the present calculations I have also used
the positional data, the NG~solutions as well as the original and
future orbits for 12 of 19~LP~comets given in the Paper~II (see
Table~1 and 2 therein).

Three samples of comets with the detectable NG effects are described
in Table~\ref{tab:3sa}, whereas the distributions of $1/a_{\rm
ori}$, $1/a_{\rm fut}$ and $\Delta E = 1/a_{\rm fut} - 1/a_{\rm
ori}$ are shown in Fig.~\ref{fig:376}.

\section{Non-gravitational effects of cometary outgassing}

The detailed discussion on modeling of the NG~effects in LP~comets
was given in Paper~II. Since one of the objectives of present
investigation is to construct the samples of homogeneously derived
NG~orbits, the standard function ${\rm g}(r)$ proposed by Marsden et
al. \cite*{MSY} was adopted here for all the comets except comet
C/1959~Y1~Burnham, for which it is evident that the asymmetric
function ${\rm g}(r(t-\tau))$ much better represents the comet
NG~motion (Paper~II). Thus, the three NG~parameters $A_{\rm 1},
A_{\rm 2}$ and $A_{\rm 3}$ were derived simultaneously with six
orbital elements from the orbital motion of each comet by assuming
that the NG~acceleration is given by equation
%\begin{equation}
$$F_{\rm i}   =  A_{\rm i} \cdot {\rm g}(r), \qquad A_{\rm i} = {\rm
~const} $$
%\end{equation}
\noindent where ${\rm i}=1,2,3$ refer to the radial, transverse and
normal components, respectively.

The investigated cometary samples I+II+III taken together contain 50
comets with determinable NG~effects. It is well known that the
actual amplitude of the NG~forces is typically at least 10 times
greater for LP~comets than for the short period comets. Using the
sample of 23 short-period comets and 7 LP~comets, Marsden et al.
(1973) have obtained that the actual magnitude of the NG~forces is
typically about $10^{-5}$ times the solar attraction at 1\,AU for
the short-period comets and (1--2)$\cdot 10^{-4}$ of the solar
attraction for the LP~comets. In Paper~II the average value of
${\big <}F_{\rm NG,max}/F_{\odot}{\big >}=1.5\cdot 10^{-4}$ was
derived for the sample of 19 LP comets containing 7 comets with $E_0
< 10^{-4}$\,AU$^{-1}$ and 12 comets with $E_0> 10^{-4}$\,AU$^{-1}$.
The results for investigated LP~comets are given in column~3 of
Table~\ref{tab:Fng}. The maxima of NG~forces, $F_{\rm NG, max}$, as
a function of the perihelion distance, $q$, are shown in
Fig.~\ref{fig:NGforce}.  The average value of the NG~forces
estimated from the whole sample of 50 comets ${\big <}F_{\rm
NG,max}/F_{\odot}{\big>}= 7.6\cdot 10^{-5}$, thus it is almost two
times lower than previously derived in Paper~II (using the sample of
19 LP~comets).

%%\begin{minipage}[t]{9.0cm}
\begin{figure}
%\vspace{0.3cm}
\begin{center}
\includegraphics[width=9.5cm]{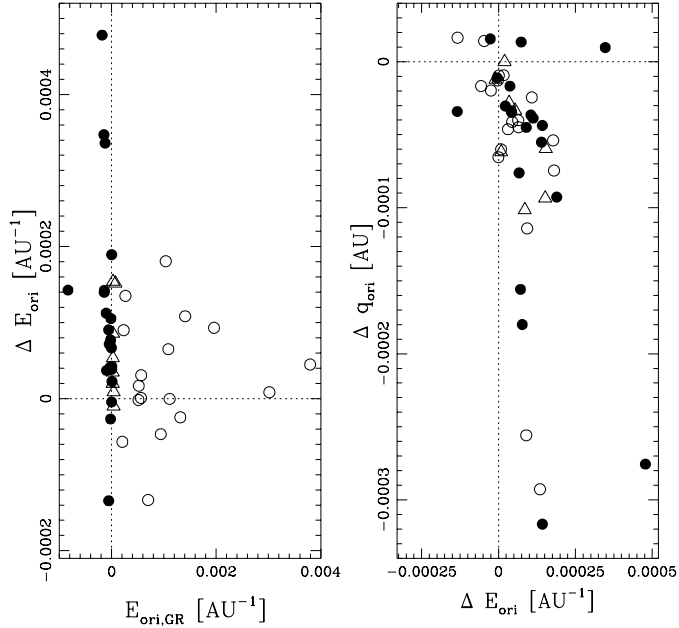}
\end{center}
\caption{{\small The original energy changes $\Delta E_{\rm ori} =
{\rm E}_{\rm ori,NG} - E_{\rm ori,GR}$ as a function of the original
energy $E_{\rm ori,NG}$ (left panel) and relation between the
perihelion distance variations, $\Delta q_{\rm ori}$ and  $\Delta
E_{\rm ori}$ (right panel). Full circles, open triangles, and open
circles show values for the sample I, II, and III, respectively.}}
\label{fig:Doria}
\end{figure}
%\end{minipage}
%\hfill

\begin{figure}
%\vspace{0.3cm}
\begin{center}
\includegraphics[width=9.5cm]{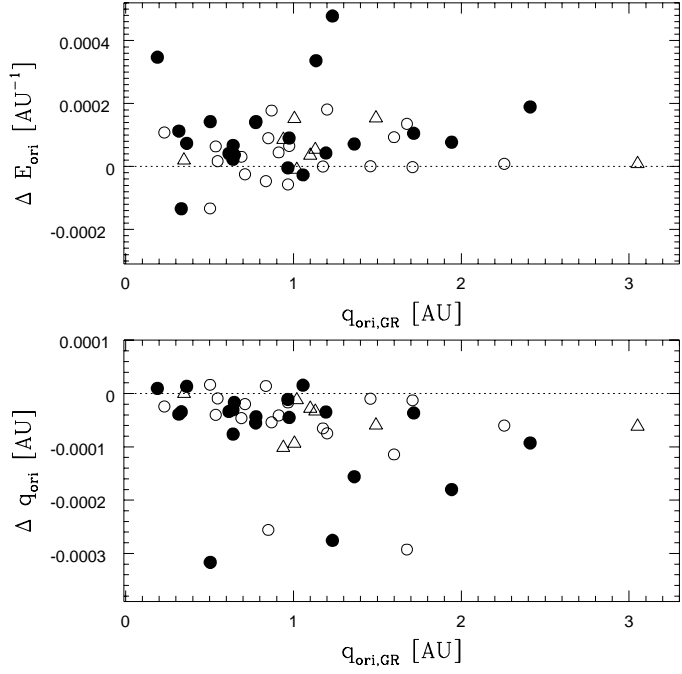}
\end{center}
\caption{{\small The original energy changes $\Delta E_{\rm ori}$
and perihelion distance variations, $\Delta q_{\rm ori}$ as a
function of the perihelion distance.  Full circles, open triangles,
and open circles  show values for the sample I, II, and III,
respectively.}} \label{fig:Doriq}
\end{figure}

\begin{figure}
%\vspace{0.3cm}
\begin{center}
\includegraphics[width=9.5cm]{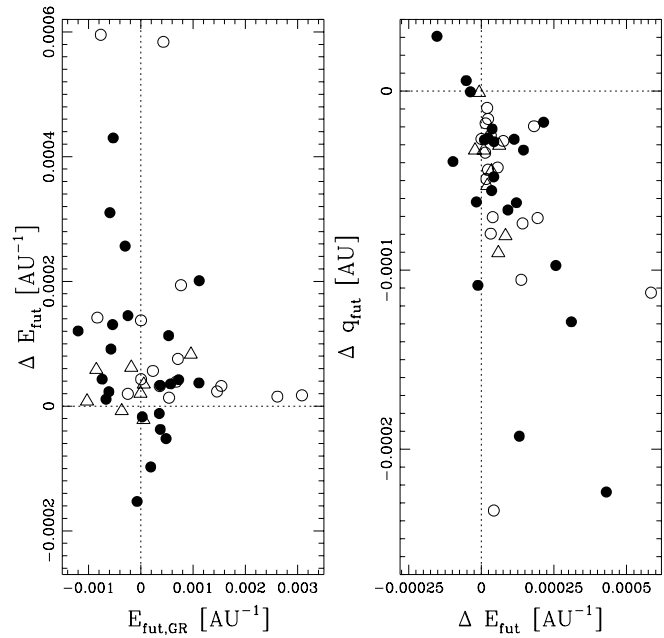}
\end{center}
\caption{{\small The same as in Fig.~\ref{fig:Doria} for future
energy changes $\Delta E_{\rm fut} = E_{\rm fut,NG} - E_{\rm
fut,GR}$ and future perihelion changes $\Delta q_{\rm fut}$}}
\label{fig:Dfuta}
\end{figure}

\begin{figure}
%\vspace{0.3cm}
\begin{center}
\includegraphics[width=9.5cm]{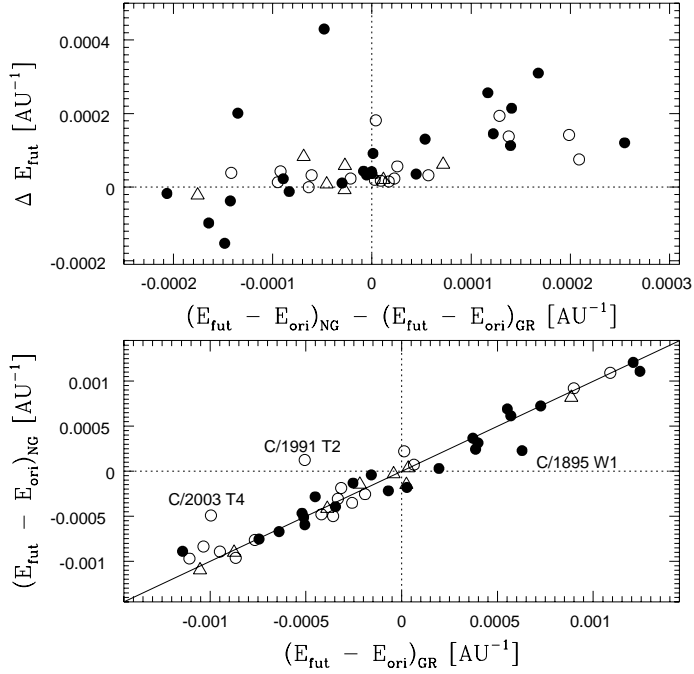}
\end{center}
\caption{{\small Bottom panel: Comparison between energy changes
during the comet crossing through the Solar System for the
non-gravitational and strictly gravitational case. Points are coded
as in Figs~\ref{fig:Doria}--\ref{fig:Dfuta}. Top panel: Changes
between the non-gravitational and strictly gravitational case shown
in the plot $\Delta (E_{\rm fut}-E_{\rm ori})$ vs $\Delta E_{\rm
fut}$.}} \label{fig:Dngr}
\end{figure}

\begin{table*}
\begin{center}
\caption{{\small Two models derived for modification of $E_{\rm
ori}={{\rm a}_{\rm ori}}^{-1}$ due to the NG effects. Model~M1 is
based on the median values of $\Delta E_{\rm ori} = E_{\rm ori,NG} -
E_{\rm ori,GR}$ and median values of $\Delta E_{\rm fut} = E_{\rm
fut,NG} - E_{\rm fut,GR}$  (columns 4--5), whereas Model~M2 -- on
the average values of ${\big <}\Delta E_{\rm ori}{\big
>}$, and ${\big <} \Delta E_{\rm fut}{\big >}$, (columns
6--7). The asterisk $^*$ denotes that two comets (C/1991~T2,
C/2003~T4) with extremely large $\Delta E_{\rm fut}$ value of about
$600\cdot 10^{-6}$\,AU$^{-1}$ were excluded. The ranges of $\Delta
E_{\rm ori}$ and $\Delta E_{\rm fut}$ are shown in columns 8 and 9,
respectively. In column~3 the mean values of the NG-effects in
perihelions, ${\big <}F_{\rm NG,max}/F_{\odot}{\big
>}$, are given.}} \label{tab:Fng} \vspace{0.10cm}
%{\setlength{\tabcolsep}{1.0mm} { \sixrm
{\small
\begin{tabular}{ccccccccc} \hline \hline
%\begin{tabular}{ccc}
%~\hspace*{5.5cm}& {\hspace*{4.0cm}}           & {\hspace*{4.0cm}}      \\
%\multicolumn{3}{r}{\bf Models used for Oort spike : }       & \multicolumn{2}{c}{\bf Model A (Median model)}& \multicolumn{2}{c}{\bf Model B (Mean model)}& \multicolumn{2}{c}{Range of}            \\
%Sample & Number     & ${\big <}F_{NG,max}/F_{\odot}{\big >}$& \multicolumn{2}{c}{median value of}           & \multicolumn{2}{c}{mean value of}           & $\Delta E_{\rm ori}$ & $\Delta q_{\rm ori}$ \\
%       & of. comets & in units of $10^{-5}$                 & $\Delta E_{\rm ori}$ &  $\Delta q_{\rm ori}$          & $\Delta E_{\rm ori}$ &  $\Delta q_{\rm ori}$        &     [AU$^{-1}$]  &     [AU]         \\
\multicolumn{3}{r}{\bf Models used for Oort spike : }       & \multicolumn{2}{c}{\bf Model M1}      & \multicolumn{2}{c}{\bf Model M2}  & &          \\
Sample & Number     & ${\big <}F_{NG,max}/F_{\odot}{\big >}$& \multicolumn{2}{c}{Median value of}   & \multicolumn{2}{c}{Mean value of} & \multicolumn{2}{c}{Range of} \\
       & of         &                                       & $\Delta E_{\rm ori}$ &  $\Delta E_{\rm fut}$ & $\Delta E_{\rm ori}$ &  $\Delta E_{\rm fut}$ & $\Delta E_{\rm ori}$ &  $\Delta E_{\rm fut}$  \\
       & comets     & in units of $10^{-5}$                 & \multicolumn{6}{c}{in units of $10^{-6}$AU$^{-1}$} \\
 \hline
% I        & 29      &  10.03  & 73.60   & 57.55     & --     &   --       & $-297$ -- $2476$ & $-255$ -- $2128$ \\
% I        & 29-2    &         & 71.44   & 43.31     & 99.63  &  94.80     & $-297$ -- $496$ & $-255$ -- $490$  \\
 I        & 23      &  9.3    & 73.60   & 42.34     & 99.54  &  81.26     & $-134$ -- $478$ & $-153$ -- $430$  \\
 II       &  8      &  5.3    & 44.22   & 27.58     & 62.30  &  20.61     &  $-10$ -- $154$ &  $-22$ -- $83$    \\
 III      & 19      &  6.9    & 30.77   & 32.39$^*$     & 39.58  &  61.54$^*$ & $-133$ -- $181$ & $0$ -- $193^*$   \\
% I        & 29      &  10.032  & 73.60   & $-39.17$  & --     &   --       &101.81 & $-74.28^*$ \\
% I        & 29-2    &          & 71.44   & $-43.47$  & 99.63  & $-96.54^*$ & 78.21 & --         \\
% II       &  8      &  5.2902  & 44.22   & $-46.43$  & 62.30  & $-48.73$   & 43.02 & $-41.59$   \\
% III      & 19      &  6.8548  & 30.77   & $-41.19$  & 39.58  & $-60.58$   & 41.00 & $-43.38$   \\
\hline
% I+II     & 37      &  8.735  & 71.44   & 43.31      & --     &  --        & $-297$ --  $2476$ & $-255$ -- $2128$ \\
% I+II     & 37-2    &         & 66.87   & 42.34      & 91.10  & 79.98      & $-297$ -- $496$ & $-255$ -- $490$  \\
 I+II     &  31      &  8.0    & 66.87   & 37.16      & 89.93  & 67.93      & $-134$ -- $478$ & $-153$ -- $430$  \\
% I+II+III & 56      &  8.046  & 60.72   & 42.71     &  --     &  --        & $-297$ -- $2476$ & $-255$ -- $2128$ \\
% I+II+III & 56-2    &         & 55.56   & 40.70     & 72.97  & 93.05      & $-297$ -- $496$ & $-255$ -- $595$  \\
%% I+II+III & {\bf 50} & {\bf 7.6}& {\bf 58.74}   & {\bf 38.11}     & {\bf 72.62}  & {\bf 86.63}      & {\bf $-134$ -- $478$} & {\bf $-153$ -- $430$}  \\
 I+II+III  & 50      &  7.6     & 58.74   &  36.41$^*$& 72.62  & 65.66$^*$  & $-134$ -- $478$ & $-153$ -- $430$
   \\
% I+II     & 37      &  8.7355  & 71.44   & $-39.17$  &        &            & 82.60 & $-64.41^*$ \\
% I+II     & 37-2    &          & 66.87   & $-43.47$  & 91.10  & $-85.29^*$ & 66.63 & --         \\
% I+II+III & 56      &  8.0457  & 60.72   & $-39.39$  &        &            & 65.96 & $-55.98^*$ \\
% I+II+III & 56-2    &          & 55.56   & $-40.64$  & 72.97  & $-74.48^*$ & 56.33 & --         \\
\hline
\end{tabular}}
%}}
\end{center}
\end{table*}

%Table~\ref{tab:new}b

\section{Variations of original and future energy due to NG effects}

Variations of the original energy, $\Delta E_{\rm ori} = 1/{\rm
a}_{\rm ori,NG} - 1/a_{\rm ori,GR}$, as a function of $1/{\rm
a}_{\rm ori,GR}$ and as a function of $\Delta q_{\rm ori} = q_{\rm
ori,NG} - q_{\rm ori,GR}$ are displayed in Fig~\ref{fig:Doria}.

Among 23~comets of the sample~I (full circles in
Fig.~\ref{fig:Doria}) four~objects have negative $\Delta E_{\rm
ori}$; one such object is in the sample~II (open triangles), and six
in the sample~III (open circles). Thus 11 of 50~comets with
detectable NG-effects have original orbits more hyperbolic than
their strictly gravitational orbits are. One can see in
Fig.~\ref{fig:Doria} that the values of $\Delta E_{\rm ori}$ are
significantly more spread for the sample~I than for the samples~II
and III. Fig.~\ref{fig:Doria} also does not reveal any significant
correlations between $\Delta E_{\rm ori}$ and $\Delta q_{\rm ori}$
(top panel). The values of $\Delta q_{\rm ori}$ span a range
$-829\cdot 10^{-6}$\,AU~$ < \Delta q_{\rm ori} < +16\cdot
10^{-6}$\,AU, excluding extremely positive value of $\Delta {\rm
q}_{\rm ori}=9860\cdot 10^{-6}$\,AU for C/1997~BA6. Only 6 of
50~comets have positive $\Delta q_{\rm ori}$. The median values of
$\Delta q_{\rm ori}$ for three basic samples are very similar and
amount to about $-40\cdot 10^{-6}$\,AU. Thus, it is evident that
nearly parabolic comets have systematically smaller $q_{\rm ori}$
and more positive $E_{\rm ori}$ due to the NG-forces. The respective
relations for the future orbits between energy and perihelion
changes, and energy and energy variations are shown in
Fig~\ref{fig:Dfuta}. The range of $\Delta E_{\rm fut}$ is similar to
the incoming orbits ($-160\cdot 10^{-6}$\,AU$^{-1}\; <\Delta E_{\rm
fut}<+600\cdot 10^{-6}$\,AU$^{-1}$) and the systematic trend to more
elliptical future NG-orbits is also clearly visible. Within the
whole sample of 50~comets only eight have $E_{\rm fut}<0$, however
six of them are new comets (about 25\% of sample~I) and remaining
two are Oort spike comets.

Deviations from the solid line given in the bottom panel of
Fig.~\ref{fig:Dngr} show how strongly the NG-effects change the
cumulative planetary perturbations during the comet's passage
through the solar system. The values of these deviations are mostly
inside the range $-300\cdot 10^{-6}$\,AU$^{-1}\; <(E_{\rm fut}-
E_{\rm ori})_{\rm NG} - (E_{\rm fut}-E_{\rm ori})_{\rm GR} <
300\cdot 10^{-6}$\,AU$^{-1}$, except three comets with labels in the
bottom panel of Fig.~\ref{fig:Dngr}. The mean value of $\mid(E_{\rm
fut}-E_{\rm ori})_{\rm NG} - (E_{\rm fut}-E_{\rm ori})_{\rm GR}\mid$
is equal to $107\cdot 10^{-6}$\,AU$^{-1}$. Since the orbital periods
of LP~comets are much longer than those of perturbing planets it is
usually assumed that LP~comets are subjected to random change of
energy during their passage through the solar system. Thus the
Gaussian distribution adequately describes the $\delta E= E_{\rm
fut}-E_{\rm ori}$ distribution. The standard deviation, $\sigma$,
calculated from the best fitted Gaussian distribution of $\delta E$
to the first quality orbits is $517\cdot 10^{-6}$\,AU$^{-1}$ (sample
A+B; Table~\ref{tab:3sa}) and is $624\cdot 10^{-6}$\,AU$^{-1}$ for
the first and second quality orbits (375 comets). Both values are
smaller than the adopted value
%This result is in a very good agreement with adopted value
of $\sigma =666\cdot 10^{-6}$\,AU$^{-1}$ given by
Yabushita~\cite*{Yab79} on the basis of various $q$~values published
earlier. Summarizing, the average error of approximately 20\% in the
value of planetary perturbations should be expected for individual
comet due to omission the NG-effects.

In Table~\ref{tab:Fng} the median and mean values of $\Delta E_{\rm
ori}$ and $\Delta E_{\rm fut}$ are given for three basic samples of
LP-comets. One can see that the median values (Model~M1 in the
table) as well as the mean values (Model~M2) are greatest for the
new comets (sample~I), as compared to the 'Oort spike' and the 'old'
comets (samples~II and III). The mean values are systematically
greater than the median values. Since $\Delta E_{\rm ori}$ as well
as $\Delta E_{\rm fut}$ have significant scatter the median values
seem to be more representative than the mean values for the typical
variations of $\Delta E_{\rm ori}$ and $\Delta E_{\rm fut}$ due to
NG-effects. The same is true for $\Delta q_{\rm ori}$ discussed
above. Corrections of $\Delta E_{\rm ori}$ given in the
Table~\ref{tab:Fng} were applied to all new+Oort spike comets in the
section~7 according to Model~M1 based on the median values as well
as Model~M2 based on the mean values.

Fig~\ref{fig:Doriq} shows the sparse distribution of the original
energy variation, $\Delta E_{\rm ori}$ in function of perihelion
distance $q$ (top panel) and similar behaviour of $\Delta q_{\rm
ori}$ versus $q$ (bottom panel). The analogical spread distributions
were noticed also for the future energy variations and future
perihelion changes in function of $q$.

Though the analysis of the mean and median values of $\Delta E_{\rm
ori}$ are given for comets with the first class quality orbits, it
is interesting to mention that among the 'new' comets with the
second class quality orbits two objects,
% Among the 'new' comets (sample I, full circles in the figure) two objects,
C/1975~X1 Sato and C/1955~O1 Honda, have very large values of
$\Delta E_{\rm ori}$, far outside the interval given in
Fig.~\ref{fig:Doria}.
%outside this figure.
Both comets have very large $1/a_{\rm ori} > 1000\cdot
10^{-6}$\,AU$^{-1}$ for the NG~orbit in comparison to their negative
value of strictly gravitational $1/a_{{\rm ori}}$. It is interesting
that both comets belong to a group of three objects (all with the
second quality class orbits) with the most negative $1/a_{\rm
ori,GR}$ (less than $-500\cdot 10^{-6}$\,AU$^{-1}$)
(Table~\ref{tab:new}). In a consequence their $\Delta E_{\rm ori}$
are also extremely large, namely $\Delta E_{\rm ori}\simeq 1860\cdot
10^{-6}$\,AU$^{-1}$ for C/1975~X1 and $\Delta E_{\rm ori}\simeq
2480\cdot 10^{-6}$\,AU$^{-1}$ for C/1955~O1, whereas the values of
$\Delta E_{\rm ori}$ of the 50~comets with the first quality orbits
are between $-290$ and $+500$ in  units of $\cdot
10^{-6}$\,AU$^{-1}$ (Fig.~\ref{fig:Doria}). Comet 1955~O1~Honda is
additionally peculiar since its two distinct nuclei were reported by
Roemer \cite*{Roe55} and by van~Biesbroeck~\cite*{Bie57}. Based on
their observations Sekanina~\cite*{Sek79} concluded that this comet
must have broken up at a large heliocentric distance long before its
discovery. The second comet, C/1975~X1 Sato, with the largest
negative $1/{\rm a}_{\rm ori,GR}$, is also peculiar due to its very
small total absolute magnitude (${\rm H}_{10} \leq 10$) \cite{Mar70}
that placed this object among the intrinsically faintest LP~comets.
However both comets have quite ordinary NG~accelerations.

\subsection{Variations of the original semimajor axes with perihelion distance}

\begin{table*}
\begin{center}
\caption{{\small The average values of the original energy, ${\big
<}E_{\rm ori}{\big >}$ of strictly gravitational orbit (column~3)
and the NG~orbit (column~5) for comets of detectable NG~effects
binned according to their $q$~values.}} \label{tab:meanew}
\vspace{0.10cm}
%{\setlength{\tabcolsep}{1.0mm} { \sixrm
{\small \begin{tabular}{cccccc} \hline \hline
\multicolumn{6}{c}{{\bf Sample I+II,  number of comets : 31} (open circles and triangles in Fig.~\ref{fig:mean1} )} \\
 && \multicolumn{2}{c}{strictly gravitational orbits} & \multicolumn{2}{c}{NG orbits} \\
range of $q$  &  number & ${\big <}1/a_{\rm ori}{\big >}$ &  ${\big <}q_{\rm ori}{\big >}$ & ${\big <}1/a_{\rm ori}{\big >}$  & ${\big <}q_{\rm ori}{\big >}$  \\
%%0.00--0.50  &  8 & -0.00008703 & 0.2843 &  -0.00000249 & 0.2843  \\
%%0.50--1.00  & 11 & -0.00011461 & 0.7675 &  -0.00004043 & 0.7674  \\
%%1.00--3.10  & 16 & -0.00001118 & 1.5985 &   0.00009481 & 1.5990  \\
%%all         & 35 & -0.00006103 & 1.0377 &   0.00003007 & 1.0371  \\
0.00--0.50  &  5 & -0.00006257 & 0.3103   &  0.00002117  & 0.3103  \\
0.50--1.00  & 10 & -0.00004255 & 0.7665   &  0.00002478  & 0.7665  \\
1.00--3.10  & 16 & -0.00001118 & 1.5985   &  0.00009481  & 1.5990  \\
all         & 31 & -0.00002959 & 1.1223   &  0.00006034  & 1.1226  \\ \hline
%&&&&& \\
\multicolumn{6}{c}{{\bf Sample III,  number of comets : 19} (open squares in Fig.~\ref{fig:mean1})} \\
 && \multicolumn{2}{c}{strictly gravitational orbits} & \multicolumn{2}{c}{NG orbits} \\
range of $q$  &  number & ${\big <}1/a_{\rm ori}{\big >}$ &  ${\big <}q_{\rm ori}{\big >}$ & ${\big <}1/a_{\rm ori}{\big >}$  & ${\big <}q_{\rm ori}{\big >}$  \\
0.00--1.00  & 12  & 0.00210388 & 0.7195 &   0.00213188 & 0.7194  \\
0.50--1.00  & 11  & 0.00216729 & 0.7638 &   0.00218800 & 0.7638  \\
1.00--2.50  &  7  & 0.00120854 & 1.5826 &   0.00126796 & 1.5825  \\
all         & 19  & 0.00177402 & 1.0374 &   0.00181360 & 1.0374  \\
\hline
\end{tabular}}
%}}
\end{center}
\end{table*}

\begin{table*}
\begin{center}
\caption{{\small The average values of the original energy, ${\big
<}E_{\rm ori}{\big >}$ for sample~A (column~3) and sample~AA
(column~5) for comets divided according to their $q$ values into 13
bins (crosses and black dots at the top panel in
Fig.~\ref{fig:mean1}).}} \label{tab:mea145} \vspace{0.10cm}
%{\setlength{\tabcolsep}{1.0mm} { \sixrm
{\small \begin{tabular}{ccccccc} \hline \hline
%\multicolumn{6}{c}{\bf Sample I+II,  number of comets : 35 (open circles and triangles in Figure 'mean_b.ps')} \\
 & \multicolumn{3}{c}{s a m p l e~~ A} & \multicolumn{3}{c}{s a m p l e~~ AA} \\
range of $q$  &  number & ${\big <}1/a_{\rm ori}{\big >}$ &  ${\big <}q_{\rm ori}{\big >}$ &  number & ${\big <}1/a_{\rm ori}{\big >}$  & ${\big <}q_{\rm ori}{\big >}$  \\
  0.00--0.50  &  9 & -0.00001379  & 0.2707 &   9 &   0.00000344 &   0.2849  \\
  0.25--0.75  & 10 &  0.00000232  & 0.4842 &  15 &   0.00000989 &   0.5039  \\
  0.50--1.00  & 10 & -0.00001252  & 0.8188 &  15 &   0.00001729 &   0.7699  \\
  0.75--1.25  & 14 &  0.00000433  & 1.0132 &  17 &   0.00002287 &   1.0232  \\
  1.00--1.50  & 18 & -0.00001080  & 1.2377 &  19 &  -0.00000352 &   1.2125  \\
  1.25--1.75  & 21 & -0.00000551  & 1.4953 &  20 &   0.00000135 &   1.4955  \\
  1.50--2.00  & 14 &  0.00003018  & 1.6694 &  15 &   0.00003906 &   1.6876  \\
  2.00--3.00  & 14 &  0.00002380  & 2.3954 &  13 &   0.00002584 &   2.3942  \\
  3.00--4.00  & 23 &  0.00003380  & 3.3684 &  23 &   0.00003567 &   3.3683  \\
  4.00--5.00  & 16 &  0.00002787  & 4.4031 &  16 &   0.00002787 &   4.4031  \\
  5.00--6.00  & 14 &  0.00004020  & 5.5000 &  14 &   0.00004020 &   5.5000  \\
  6.00--7.00  &  6 &  0.00002835  & 6.3234 &   6 &   0.00002835 &   6.3234  \\
  7.00--10.00 &  7 &  0.00004073  & 8.1556 &   7 &   0.00004073 &   8.1556  \\
  \hline
  0.00--0.50  & 37 & -0.00001199  & 0.8893 &  43 &   0.00000519 &   0.8640  \\
  0.50--1.00  & 94 &  0.00003188  & 4.0091 &  94 &   0.00003411 &   4.0041  \\
  all         &131 &  0.00001949  & 3.1279 & 137 &   0.00002503 &   3.0185  \\
  \hline
&
%  1.75--2.50  &  15  &  0.00001602  &  2.1297  &  &  & \\
%  2.50--3.50  &  20  &  0.00003720  &  3.0562  &  &  & \\
%  3.50--4.50  &  18  &  0.00002513  &  4.0100  &  &  & \\
%  4.50--5.50  &   9  &  0.00002998  &  5.0084  &  &  & \\
%  5.50--6.50  &  11  &  0.00003656  &  5.9402  &  &  & \\
%  6.50--11.5  &   7  &  0.00004342  &  8.3429  &  &  & \\
% \hline
\end{tabular}}
%}}
\end{center}
\end{table*}

\begin{figure}
%\vspace{0.3cm}
\begin{center}
\includegraphics[width=10.0cm]{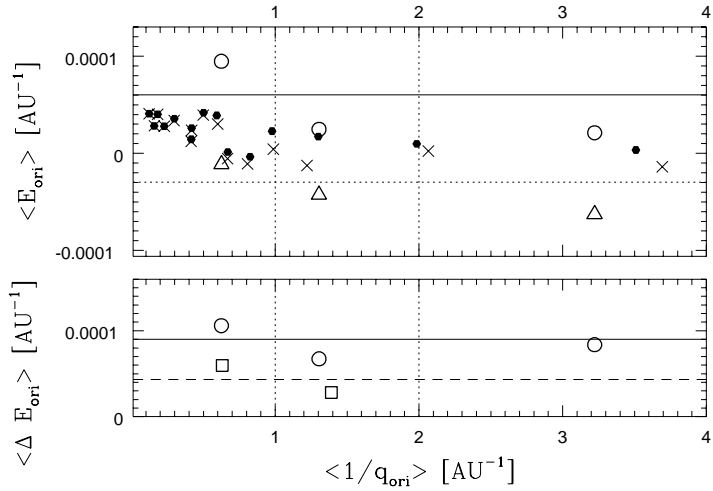}
\end{center}
\caption{{\small The average values of the original energy, ${\big
<}E_{\rm ori}{\big >}$ (top panel) and the average value of the
difference between the original energy of strictly gravitational
orbit and the NG~orbit ${\big <}\Delta E_{\rm ori}{\big >}$ (bottom
panel) for comets divided according to their $q$ values. Open
circles and triangles correspond to the 'new' $+$~Oort spike comets
(sample~I$+$II), whereas the open squares represent the 'old' comets
(sample~III; Table~\ref{tab:meanew}). Crosses in the top panel
represent the mean values of  ${\big <}E_{\rm ori}{\big >}$ for the
sample of 131 comets of ${\big <}E_{\rm ori}{\big
>}< 10^{-4}$\,AU$^{-1}$ (sample~A, Table~\ref{tab:3sa}) divided into 13
bins (Table~\ref{tab:mea145}), whereas the black dots show the same
quantities for the sample of 137 comets (sample~AA,
Table~\ref{tab:mea145})}} \label{fig:mean1}
\end{figure}

Marsden et al.~\cite*{MSY} have shown that the mean of $1/{\rm
a}_{\rm ori}$ of 'new' comets tend to be negative at small
perihelion distance. They examined the sample of 61 class~I comets
(i.e. comets with well-determined orbits), and concluded that
presumably this is due to the influence of the NG~effects.

Here, I examine this problem on the basis of 56~comets with the
detectable NG~effects. To determine the average values of the
reciprocals of the original semimajor axes, $E_{\rm ori, NG} =
1/a_{\rm ori,NG}$ and $E_{\rm ori, GR} = 1/a_{\rm ori,GR}$, the two
subgroups were constructed, first consists of objects with $1/{\rm
a}_{\rm ori} < 10^{-4}$\,AU$^{-1}$ (sample I+II) and second --
objects with $1/a_{\rm ori} > 10^{-4}$\,AU$^{-1}$ (sample III).
% The two comets (C/1975~X1 and C/1955~O1 mentioned before) with extremely
%large $\Delta {\rm E}_{\rm ori}$ for analysis of the mean values of
%$1/a_{\rm ori}$ and $\Delta E_{\rm ori}$ were excluded.
The sample~I+II was divided onto three subgroups containing 5~comets
($q < 0.5$\,AU), 10~comets (0.5\,AU~$ < q < 1.0$\,AU) and 16~comets
($q > 1.0$\,AU), respectively.

\noindent Sample~III of 'old' comets contains only one object of $q
<0.5$\,AU. So it was practical to divide this sample on two
subgroups only, with $q < 1.0$\,AU (12 objects) and $q
> 1.0$\,AU (7 objects), respectively.

The average values of the original energy, ${\big <}E_{\rm ori,
NG}{\big >}$ and ${\big <}E_{\rm ori, GR}{\big >}$ for comets
divided according to their $q$ values are given in
Table~\ref{tab:meanew}, whereas the top panel of
Fig.~\ref{fig:mean1} shows these average values versus the average
$1/q_{\rm ori}$. These values based on the samples of comets with
the detectable NG~effects could be compared with the complete sample
of 131~comets of $E_{\rm ori}< 10^{-4}$\,AU$^{-1}$ extracted from
the MW~Catalogue. Table~\ref{tab:mea145} gives the average values of
${\big <}E_{\rm ori, GR}{\big
>}$ for 13 overlapping groups of the sample~A, and the crosses in the
top panel of Fig.~\ref{fig:mean1} show their distribution vs.
reciprocals of $q$.

The impact of the NG~effects on the $E_{\rm ori}$ of comets in the
sample~A is now assessed. First, the NG~orbits derived for 31~comets
of sample I+II were inserted instead of the strictly gravitational
orbits taken from the MW~Catalogue. Next, 11 gravitational first
class orbits of comets with the indeterminable NG~effects were taken
instead the catalogue orbits. Thus, the new orbits were implemented
for 42 comets, where 13~of them belonged to the comets outside the
sample~A (originally the comets having NG~orbits in the MW~Catalogue
were ejected from the sample~A). On the other hand, seven comets
belonging to the sample~A have NG~orbits with $E_{\rm ori, NG}$
greater then $10^{-4}$\,AU$^{-1}$. Finally, the new corrected sample
of comets with $E_{\rm ori}< 10^{-4}$\,AU$^{-1}$ (designed as the
sample~AA; Table~\ref{tab:3sa}) contains 137 objects, including 27
with the NG~orbits. The average values of ${\big <}E_{\rm ori}{\big
>}$ for thirteen overlapping groups of this sample are given in the
column~6 of Table~\ref{tab:mea145}, and are shown as black dots at
the top panel of Fig.~\ref{fig:mean1}. One can see that, the average
values of $E_{\rm ori}$ have smaller dispersion than the respective
values in the sample~A. These dots could be compared to the open
triangles inside the same $q$~ranges. Among the sample~I+II 74\% of
objects have $E_{\rm ori,GR} < 10^{-5}$\,AU$^{-1}$, whereas the
sample~AA includes less than 25\% of such comets. In a consequence,
the triangles are situated below the respective dots. The behaviour
of ${\big <}E_{\rm ori}{\big >}$ as a function of ${\big <}1/q_{\rm
ori}{\big >}$ differs from the linear relation derived for 61 comets
of class~I by Marsden et al.~\cite*{MSY}.

It is clearly visible that the average $E_{\rm ori}$ in the
$1/q$~intervals for $q<1.5$\,AU are systematically smaller than for
$q>1.5$\,AU. The original energies ${\big <}E_{\rm ori}{\big
>}$ for the low~$q$ orbits cluster around $+5\cdot
10^{-6}$\,AU$^{-1}$, while the large~q orbits concentrate around
$+34\cdot 10^{-6}$\,AU$^{-1}$ (Table~\ref{tab:mea145},
Fig.~\ref{fig:mean1}). The difference of $\sim 40\cdot
10^{-6}$\,AU$^{-1}$ between ${\big <}E_{\rm ori}{\big>}$ for these
two intervals of $q$ is smaller than the mean changes of the
original energy between the NG-orbit and strictly gravitational
orbit ${\big <}\Delta E_{\rm ori}{\big >}$ represented by open
circles and solid horizontal line in the bottom panel of
Fig.~\ref{fig:mean1}. Thus the tendency of the negative mean value
of $E_{\rm ori}$ at small~$q$ for the sample of the 'new' and Oort
spike comets (sample~AA) would be canceled by the correct
modification of known gravitational orbit due to NG-forces.

\section{The Oort spike modification due to NG~force}

\begin{figure}
%\vspace{0.3cm}
%\includegraphics[width=9.0cm]{all_158a.eps}
\begin{center}
\includegraphics[width=12.0cm]{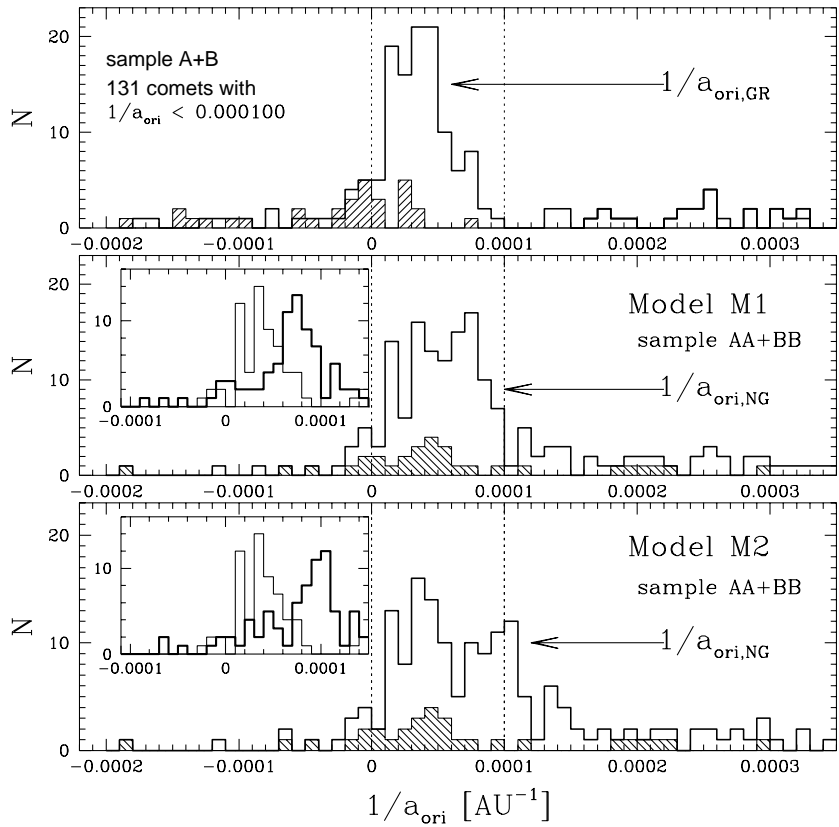}
\end{center}
\caption{{\small Top panel:  The strictly gravitational Oort spike
distribution, $E_{\rm ori,GR}$ for the sample~A (solid line
histogram) and the distribution of $E_{\rm ori,GR}$ for the basic
sample~I+II (shaded histogram). Middle and bottom panels: The Oort
spike distributions (thick solid lines) corrected for the NG-effects
according to Model~M1 and Model~M2, respectively, where corrections
were applied to comets with q$<3.1$\,AU. The shaded areas at both
lower panels represent the distribution of $E_{\rm ori,NG}$ for the
basic sample I+II. The insets show the respective distributions for
comets with q$<3.1$\,AU (thick solid line), and comets with
q$>3.1$\,AU (thin solid line), separately.}} \label{fig:spike}
\end{figure}

\begin{figure}
\begin{center}
\includegraphics[width=10.0cm]{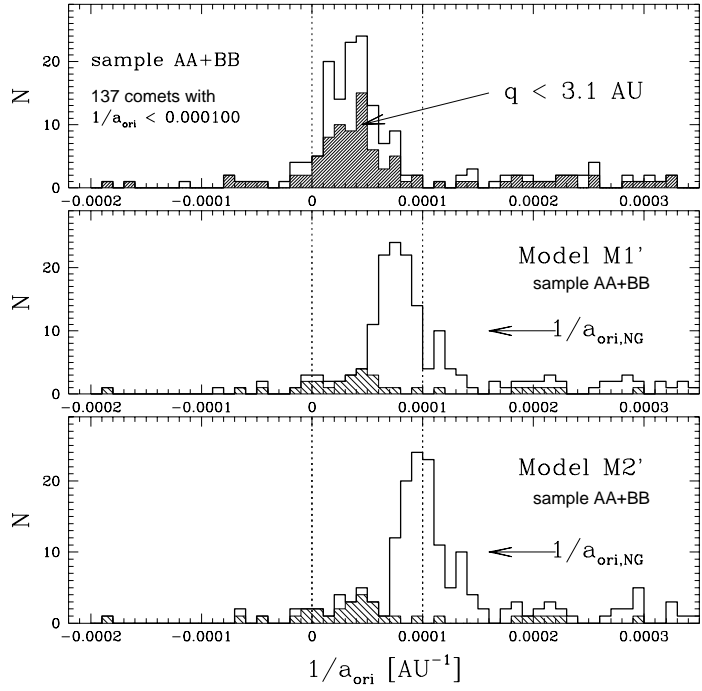}
\end{center}
\caption{{\small Top panel: The Oort spike distribution of cometary
energies for the sample~AA (thick solid line), where the shaded
distribution represents the comets with $q<3.1$\,AU. Middle and
bottom panels: the same as in Fig.~\ref{fig:spike} where all the
comets of sample~AA were corrected for the NG-effects according to
Model~M1 and Model~M2, respectively. }} \label{fig:spike2}
\end{figure}

\begin{figure}
%\vspace{0.3cm}
\begin{center}
\includegraphics[width=10.5cm]{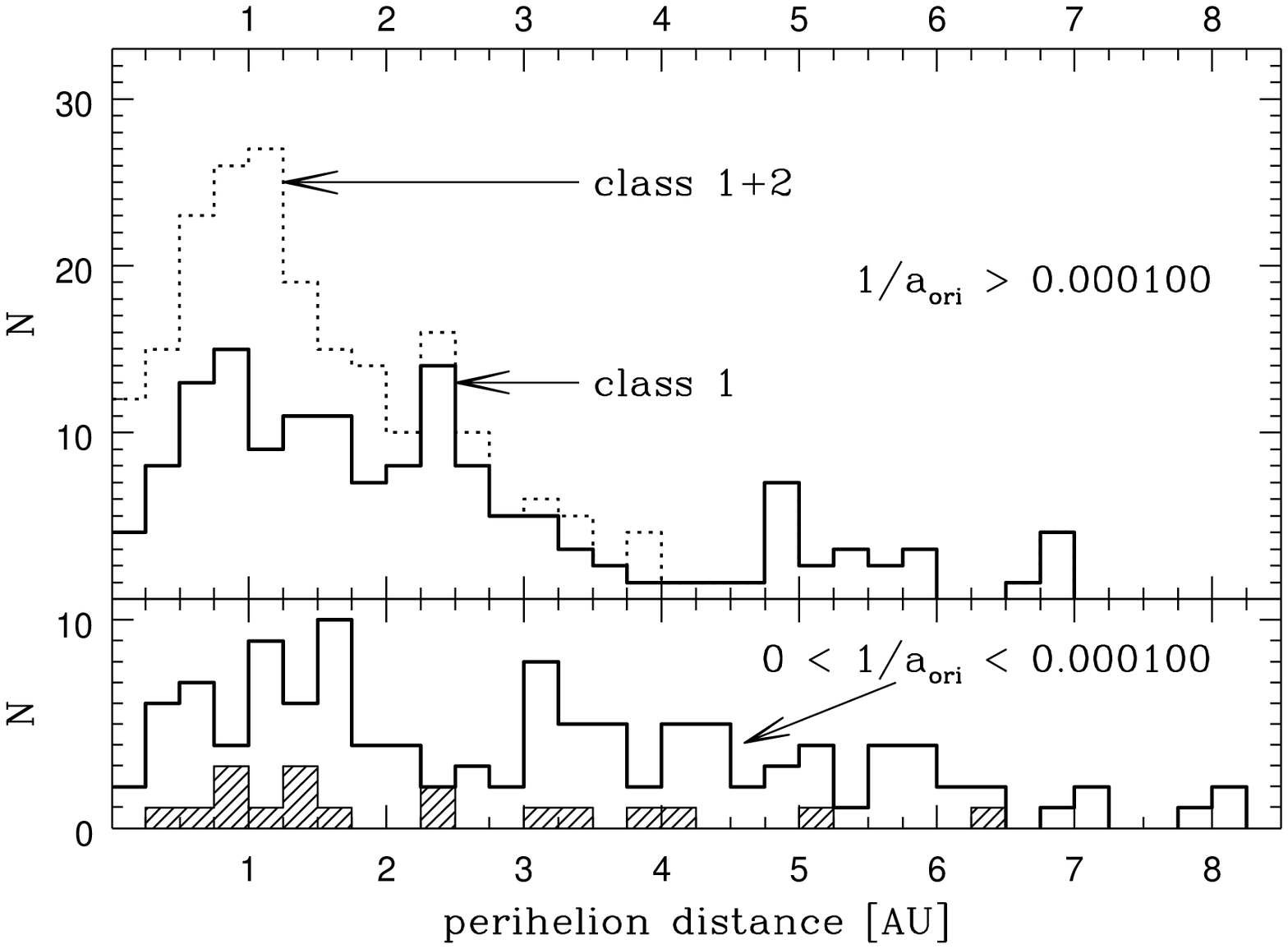}
\end{center}
\caption{{\small Comparison between distributions of perihelions of
'old' comets collected in sample~BB (top panel, solid line
histogram) and comets of sample~AA (bottom panel). The sample~AA is
divided into hyperbolic (shaded histogram), and new + Oort~spike
comets (solid line distribution). The dotted line histogram in the
top panel represents the distribution of comets with orbits of
quality classes 1 and 2.}
%The differences in the galactic coordinates between the perihelion
%direction distributions of the 'old' comets (thin solid line) and
%the Oort spike comets (thick solid line) are presented in the two
%bottom panels. The influence of the Galactic tide on the Oort spike
%comet distribution as a function of galactic latitude is clearly
%visible
} \label{fig:qdist}
\end{figure}

The sample~AA of 137~comets contains 27 NG-orbits. Among the
remaining 110 objects, 53 comets have perihelion distance
$q>3.1$\,AU. This value of $q$ was assumed as the limiting distance
for activity of the NG~forces. Thus, for 68~comets with strictly
gravitational orbits and $q<3.1$AU the $1/a$-corrections given in
the Table~\ref{tab:meanew} according to Model~M1 and Model~M2 were
applied. The resulting distributions of $E_{\rm ori}$ are given in
the middle and bottom panel of Fig.~\ref{fig:spike}, respectively.

I first discuss the model~M2 based on mean values of $\Delta {\rm
E}_{\rm ori}$ that gives greater corrections to the original
semimajor axes. The Oort spike in the model~M2 splits in two
populations.  This is because the $1/a$-corrections have been
applied only for $q<3.1$\,AU. The left peak is centered around the
value of $E_{\rm ori}\sim 3-4\cdot 10^{-5}$AU$^{-1}$, slightly to
the right from the old position of the observed Oort spike for
strictly gravitational original energy. The newly formed right
maximum peaks around the value $E_{\rm ori}\sim 9\cdot
10^{-5}$AU$^{-1}$. The first is formed mostly by LP comets with
great perihelion distances, whereas the latter -- by comets with $q$
below 3.1\,AU. Almost 60\,\% of comets with $q > 3.1$\,AU have the
original energy in the range $0 < E_{\rm ori}<7\cdot
10^{-5}$AU$^{-1}$, whereas only 5\,\% of comets with $q
>3.1$AU are in the range $7\cdot 10^{-5}$AU$^{-1} < E_{\rm
ori}<12\cdot 10^{-4}$AU$^{-1}$ (see the inset in the bottom panel of
Fig.~\ref{fig:spike}).  This double-peaked distribution for comets
with $E_{\rm ori} < 10^{-4}$AU$^{-1}$ seems very attractive for
theoretical modeling of long-term dynamical evolution of nearly
parabolic comets.

Probably more realistic and more conservative seems Model~M1 based
on the median values of $\Delta E_{\rm ori}$. In this model the Oort
spike is transformed into significantly more spread and less peaked
distribution that is well represented by a gaussian distribution
with the maximum at $E_{\rm ori} = 5.5 \cdot 10^{-5}$AU$^{-1}$ and
$\sigma = 3.8 \cdot 10^{-5}$AU$^{-1}$. Though the whole sample
exhibits the normal distribution of $E_{\rm ori}$, some traces of
two populations are revealed when distributions of $1/a$ are plotted
separately for comets with small and large $q$. Similarly to
Model~M2, also here the comets with large $q$ dominate on the
left-half of the range $0 < {\rm E}_{\rm ori} < 10^{-4}$AU$^{-1}$,
while the comets with small $q$ are grouped mostly in the right-half
of the same range. Inset given in the middle panel of
Fig.~\ref{fig:spike} shows the respective histograms for $q<3.1$\,AU
(thick solid line) and $q>3.1$\,AU (thin solid line).

Modifications of the Oort spike generated by the improved
$1/a$-corrections applied to all the comets are presented in the
Fig.~\ref{fig:spike2}. The Oort spike is now shifted towards more
positive energies than in the Fig.~\ref{fig:spike}. The mean value
of $E_{\rm ori}$ for the Oort spike is now centered roughly on ${\rm
E}_{\rm ori} \simeq 7 - 8\cdot 10^{-5}$\,AU$^{-1}$ (middle panel)
for the median corrections and $E_{\rm ori} \simeq 9 - 10 \cdot
10^{-5}$\,AU$^{-1}$ for the mean corrections (bottom panel).

%The top panel shows the $E_{\bf ori}$ distribution for the
%sample~AA.

%Figure~\ref{fig:qdist} shows that the q-distributions of LP~comets
%with $E_{\rm ori}<10^{-4}$ (sample~AA) and $E_{\rm
%ori}>10^{-4}$ (sample~BB) are substantially different.

\section{Perihelion distance distributions of dynamically new and old comets}

The problem of the actual perihelion distributions of the LP~comets
is complex due to various observational biases. The most severe
effects are present in the old samples of LP~comets because the data
are highly incomplete beyond 1.5--2.5\,AU. Recently, Hughes
\cite*{Hug01} has shown that old discoveries of the LP~comets
(before $\sim$1900) give a maximum around 0.8\,AU. He derived that
the maximum shifts to $\sim 1$\,AU for comets discovered between
1987-1946, whereas the distribution of the LP~comets detected after
1950 is significantly broader and peaks between 1--2\,AU. He also
concluded that modern cometary catalogues of the LP~comets are still
incomplete beyond 2.5\,AU. To derive the true perihelion
distribution for the active LP~comets at large $q$ it is necessary
to introduce the corrections for the observational bias in the comet
discoveries. At present such corrections are not available and the
distributions discussed here are also subject to this bias.

The upper panel in Fig.~\ref{fig:qdist} shows the observed
perihelion distributions of the 'old' comets of class 1 (sample~BB,
solid line histogram) and class 1+2 (dotted line histogram),
separately. The observed perihelion distributions of
'new'\,$+$\,Oort spike comets (sample~AA without hyperbolic comets)
is presented in the bottom panel of Fig.~\ref{fig:qdist}. One can
see the significant differences between both histograms of comets
with the first class orbit. It is well-known that the perihelion
distribution of the LP~comets has a maximum near $q\approx 1$\,AU,
This also results from the observational selection. The theoretical
modeling of the cometary dynamics suggests that the actual number of
comets per unit perihelion distance increases with increasing $q$
\cite{Weis85}. However, probability of the comet detection
diminishes with the increasing distance from the Sun and Earth. In
effect, the observed distribution of the LP~comets is peaked near
$q=1$\,AU from the Sun. The distribution of the 'old' LP~comets
(class 1+2, dotted line histogram in the top panel of
Fig.~\ref{fig:qdist}) reflects exactly the same effect.

In contrast to the 'old' LP~comets, the perihelion distributions of
the 'new' and Oort spike comets with $0 < E_{\rm ori}<1\cdot
10^{-4}$AU$^{-1}$ are quite different (thick-line histogram in the
bottom panel of Fig.~\ref{fig:qdist}).  The numbers of objects in
the consecutive perihelion distance bins are 19 (0--1\,AU), 29
(1--2\,AU), 11 (2--3\,AU), 20 (3--4\,AU), 15 (4--5\,AU), and 25 for
$q>5$\,AU (21\%). This gives 50\% of objects with perihelions inside
sphere of 3\,AU for the new and Oort spike comets, while fraction of
the 'old' comets with $q<3$\,AU reaches 76\%. A relative excess of
the 'new' and Oort spike comets with the large perihelion distance
over the 'old' comets result -- at least to some extent -- from the
efficient processes of the gradual decline of activity during the
cometary evolutions. It is well known that to obtain the agreement
between the observed distribution of original cometary energy and
the expected one by theory some fading model must be assumed
\cite{Weis80,Ba84,WT99}. However, such modeling of fading is
complicated by a fact that
%number of 'new' and Oort spike comets with large perihelions in
%comparison to the number of 'old' comets is caused -- at least
%partially -- by differences in the comet brightness.
dynamically new comets are often anomalously bright at large
perihelion distances (especially on the inbound leg of their
orbits), probably due to sublimation of more volatile ices than
water (and possibly larger sizes of the cometary nuclei). The
observed $q$~distribution of the 'new' and Oort spike comets
(thick-line histogram in the bottom panel of Fig.~\ref{fig:qdist})
could be represented either by a single-peaked distribution with
broad maximum somewhere between 1--4\,AU or by double-peaked
distribution consisting of two populations, where the left-hand peak
has the same interpretation as in the case of the entire sample of
'LP'~comets. These comets sublimate mostly the water ice. A merged
sample of hyperbolic comets (shaded histogram in the bottom panel of
Fig.~\ref{fig:qdist}) and the 'new' and Oort spike comets (thick
line histogram) has a broad main maximum centered at 1\,AU.
Nevertheless, the local maximum for perihelions greater than 3\,AU
or local minimum within $1.75$\,AU\,$<q<3.00$\,AU is still
perceptible. Orbits within this local minimum are mostly retrograde
and this fact will be discussed in the next section.

%This interpretation is supported by Emel'yanenko and
%Bailey~(\cite{EBa98}). They have found that the maximum probability
%for transfer from the near-parabolic comet into Halley-type orbits
%occurs for $q<4$\,AU. They show that about 1\% of the Oort Cloud
%comets can survive the capture into observed population of Halley
%Type Comets (HTCs). According to their calculations the Halley
%capture probability is an order of magnitude lower ($\sim 0.2$\%) in
%the region $q>4$\,AU.

\section{Distributions of prograde and retrograde orbits}

Fernandez \cite*{Fer02} divided the observed population of the
LP~comets into 8 subgroups according to the original orbital energy
$E_{\rm ori}$. He found that distributions of Oort spike comets with
$23.8\cdot 10^{-6}$\,AU$^{-1} < E_{\rm ori}<100\cdot
10^{-6}$AU$^{-1}$ (three groups) show excess of the retrograde
orbits, while the remaining groups show a weak excess of direct
orbits. He argued that this excess of retrograde orbits is connected
with the higher probability of diffusing through the Jupiter-Saturn
barrier for comets in the retrograde orbits than for comets in the
direct orbits. I have performed a similar analysis for the corrected
global samples AA and BB dividing the first of them into two
subgroups and the second - into 3 subgroups. Within two subgroups of
the 'new'\,$+$\,Oort spike comets no excess has been obtained. The
whole sample~AA contains almost the same number of comets on direct
orbits (71 objects) as on retrograde orbits (65 objects). A small
excess of retrograde orbits was obtained for the sample of all the
'old' comets (sample~BB): 82 of 149 move on retrograde orbits.
However, among three subgroups of the 'old' comets only one of them,
defined by $100\cdot 10^{-6}$\,AU$^{-1} < E_{\rm ori}<500\cdot
10^{-6}$AU$^{-1}$, shows statistically significant excess of the
retrograde orbits (34 comets) in comparison to the direct orbits (12
comets). This sub-sample corresponds to the sub-sample~'e'
constructed by Fernandez~\cite*{Fer02}, for which the most prominent
excess of retrograde orbits is visible in the Fig.~1 of his paper.
Very similar results were derived when the sample of 286 comets with
the first class orbits (sample AA+BB) were supplemented by 113
comets with the second class orbits. Then, like previously, among
five sub-samples only this define by $100\cdot 10^{-6}$\,AU$^{-1} <
E_{\rm ori}<500\cdot 10^{-6}$AU$^{-1}$, shows statistically
significant excess of the retrograde orbits (41~comets) in
comparison to the direct orbits (16 comets).

The significant excess of the retrograde orbits was also found for
the LP~comets having 1.75\,AU\,$<q<3.00$\,AU. Table~\ref{tab:incl}
shows that such excess is evident in both corrected samples of
LP~comets. However, among the 'new' and Oort spike comets
(sample~AA) the number of the retrograde orbits exceeds four times
the number of the direct orbits, whereas within the old comets --
the excess amounts to 50\%. One should note, that among ten
dynamically new comets having perihelions between 1.75\,AU and
2.25\,AU  all move on the retrograde orbits. In the $q$~range of
1.75\,AU\,$<q<3.00$\,AU the local deficiency of the 'new'\,$+$\,Oort
spike comets is observed (see Fig.~\ref{fig:qdist}), while the
perihelion distribution of the 'old' comets has local maximum at
2.25\,AU\,$<q<2.75$\,AU. It seems that some mechanism prevents the
'new' and Oort spike comets entering the solar system on the direct
orbits to reach this range of the perihelion distances. Apparently
this bias does not affect strongly the retrograde orbits. It is
likely that the 'old' comets are also subject to the same mechanism.

%First was derived within the sub-sample of $500\cdot
%10^{-6}$\,AU$^{-1} < E_{\rm ori}<1000\cdot 10^{-6}$AU$^{-1}$
%where 49 of 77 objects (64\%) have orbits inclined to ecliptic
%more than 54\degr , whereas whole sample~BB 143 of 250 (57\%)
%have orbits inclined to ecliptic more than 54\degr.

\begin{table*}
\begin{center}
\caption{{\small Distributions of direct and retrograde orbits for
three samples of the LP~comets with perihelions between 1.75\,AU and
3.00\,AU}} \label{tab:incl} \vspace{0.10cm}
%{\setlength{\tabcolsep}{1.0mm} { \sixrm
{\small \begin{tabular}{ccccccc} \hline \hline
            & \multicolumn{2}{c}{\bf Sample AA} &  \multicolumn{2}{c}{\bf Sample BB} & \multicolumn{2}{c}{\bf Sample AA+BB} \\
range of $q$ & $i<90$\degr & $i>90$\degr & $i<90$\degr & $i>90$\degr & $i<90$\degr & $i>90$\degr  \\
1.75--2.25   &  0          &         8   &  7          &        7   &   7          &       15    \\
2.25--3.00   &  3          &         6   & 10          &       18   &  13          &       24    \\
1.75--3.00   &  3          &        14   & 17          &       25   &  20          &       39    \\
 \hline
\end{tabular}}
%}}
\end{center}
\end{table*}

\section{Summary and concluding remarks}

In the present study the NG~effects are derived for 50~LP~comets,
where 31 of them have $E_{\rm ori, GR} < 10^{-4}$\,AU$^{-1}$. It was
confirmed that NG~forces make 'hiperbolic' and elliptical osculating
orbits less eccentric than by assuming purely gravitational motion.
Such number of comets with the detectable NG~effects allow us to
give some estimates of the average value of $\Delta E_{\rm ori} =
1/a_{\rm ori, NG} - 1/a_{\rm ori, GR}$ for three subsamples of the
LP~comets. Due to large dispersion of the $\Delta E_{\rm ori}$
distributions (within each subsample) the median values of $\Delta
E_{\rm ori}$ seem to represent more adequately typical corrections
for $E_{\rm ori, GR}$ of the LP~comets with the non-detectable
NG~effects.

Applying these corrections for the all LP~comets, the estimates of
the Oort spike shift from its classical position (i.e. position
derived on the basis of pure gravitational model of comet's motion)
have been obtained.

Unfortunately, the available data warrant only simplified type of
models of the $1/a$~corrections. In the first type it was assumed
that the NG~effect are present only for $q<3.1$\,AU
(Fig.~\ref{fig:spike}), while in the second type no upper limit for
$q$ was imposed (Fig.~\ref{fig:spike2}). Although one could expect
smooth decline of the NG~effects with the increase of the perihelion
distance, no such relationship is implied by the observational data.
In particular it is not present in the our basic sample of 50~comets
(the bottom panel in Fig.~\ref{fig:mean1} and section 6.1). On the
other hand, the active comets at large perihelion distances have
been observed (e.g. C/2004~T3~Siding Spring with $q=8.9$\,AU,
C/2000~A1~Montani with $q=9.7$\,AU, and C/2003~A2~Gleason with
$q=11.4$\,AU). In such cases another formula for the NG~acceleration
due to sublimation of more volatile ices is necessary. Such modeling
of the NG~effects in the LP~comets with large perihelion distance
based on Yabushita CO-function \cite{Yab96} was performed in
Paper~II. However, no reliable NG~effects have been found for comets
with the perihelion distance greater than 3.5\,AU.

Using the model of the NG~effects limited to comets with $q\leq
3.1$\,AU, the Oort spike is transformed into significantly more
spread and less peaked distribution with the maximum at $E_{\rm ori,
NG} = 5.5 \cdot 10^{-5}$AU$^{-1}$, whereas if no upper limit on $q$
is imposed the Oort spike peaks near $E_{\rm ori, NG} \simeq
8.5\cdot 10^{-5}$\,AU$^{-1}$ (middle panels of Figs.~\ref{fig:spike}
and \ref{fig:spike2}). Assuming conservatively $E_{\rm ori, NG} =
5.5 \cdot 10^{-5}$AU$^{-1}$ the outer Oort Cloud distance from the
Sun is approximately 37\,000\,AU. Consequently the inner edge of the
Oort Cloud is situated at a distance significantly smaller than the
'classical' 20\,000\,AU assumed in the literature. The middle panel
of Fig.~\ref{fig:spike} suggests that this inner edge could be less
spectacular than in the classical point of view, i.e. in the
strictly gravitational case (top panel of Fig.~\ref{fig:spike}).
%suggested by simulations of dynamical evolution.

On the other hand, the theoretical simulations predict that the
energies of new comets should peak near $E_{\rm ori}\approx 2.9
\cdot 10^{-5}$AU$^{-1}$, i.e. at a semimajor axis of about
34\,000\,AU for the currently accepted value of a local mass density
of 0.1~M$_{\odot}$/pc$^3$ \cite{Lev01}. Assuming that Sun formed in
a denser galactic environment than it occupies now, it is possible
to obtained the Oort Cloud bound more tightly \cite{Egg99,FeB00}.
For example, taking an initial local density of 100 times greater
than the current local density, and assuming that the early Sun
spent in such environment first 20~m.y. yrs, Eggers \cite*{Egg99}
modeled the Oort cloud formation. She has found that the tightly
bound cloud of comets might have been formed at a heliocentric
distance of about 6--7~thousand\,AU.

It is obvious that such significant reduction of the Oort cloud
distance as postulated in the present paper has a deep consequences
in many aspects of theoretical modeling of the Oort Cloud formation
as well as simulations of a long-term evolution of the Oort Cloud
comets, and cometary fading and disruption. In the last decades, all
these three blocks of issues have been intensively explored (see
reviews by Dones et al. \cite*{DWLD04}, and Rickman~\cite*{Ric04}).
However, evolutionary correlations between the observed cometary
populations and their reservoirs are still incomplete and are
inconsistent in many aspects \cite{EBa98,NVZR,Ric05}
%(Emel'yanenko \& Bailey~\cite{EBa98},Nurmi et al. \cite{NVZR}, Rickman~\cite{Ric05}).
One of the most important problems raised by the recent simulations
%(Levison et al. \cite{Lev01}, Duncan et al. \cite {DQT87}, Dones et al. \cite{DWLD04})
\cite{Lev01,DQT87,DWLD04} are correlations between the outer and
inner Oort Cloud populations and their characteristics. Still more
realistic simulations are needed.

%\end{document}

%\vspace{1.0cm}

\newpage

\begin{changemargin}{-1.5cm}{-1cm}

%%%%\begin{center}
{\small \tablefirsthead{%
%  \hline
\hline \hline
  Name    &  q       & Observational  &    Number  &class& {\bf Sample} &  \multicolumn{2}{c}{$1/a_{\rm ori}$}             & \multicolumn{2}{c}{$1/a_{\bf fut}$} & Remarks \\
          &  [AU]    & interval       & of obs.    & MW  &              &  \multicolumn{4}{c}{[in units of $10^{6}$AU$^{-1}$]}   & \\
          &          &                &            & Cat.&              & grav      & NG        & grav        & NG               & \\
\hline
 \multicolumn{10}{c}{\bf (i) ~~Long period comets of $1/a_{\rm ori} < 10^{-5}$AU$^{-1}$ obtained using the MW~Catalogue orbits}   \\
  }}
{\small \tablehead{%
\hline
  Name    &  q       & Observational  &    Number  &class& {\bf Sample} &  \multicolumn{2}{c}{$1/a_{\rm ori}$}             & \multicolumn{2}{c}{$1/a_{\bf fut}$} & Remarks\\
          &  [AU]    & interval       & of obs.    & MW  &              &  \multicolumn{4}{c}{[in units of $10^{6}$AU$^{-1}$]}   & \\
          &          &                &            & Cat.&              & grav      & NG        & grav        & NG               & \\
\hline}}
\tabletail{%
\hline
\multicolumn{11}{r}{\small\sl continued on next page}\\
\hline} \tablelasttail{\hline} \tablecaption{{\small Original and
future reciprocals of semimajor axes derived in the present
investigations for the sample of comets with $1/a_{\rm ori}$
(obtained using the catalogue orbits): below $10^{-5}$\,AU (part (i)
\& (ii) of this table), $10^{-5}$\,AU\,$< 1/a_{\rm ori} <
10^{-4}$\,AU (part (iii)), and $1/a_{\rm ori} > 10^{-4}$\,AU (part
(iv)) In the columns 2--6 the perihelion distance, the observational
arc and number of observations taken for orbit determinations,
orbit's quality class and the sample designation are given. Remarks
in the last column denote: pp -- present paper; PIgr -- only
strictly gravitational model in Paper~I; PIdif -- different
selection than in Paper~I; PImore -- more observations than in
Paper~I; PaperI, PaperII -- solutions taken from Paper~I and
Paper~II, respectively.
%The asterisk $^*$ denotes that the quality class (column 4) was taken from the current edition of MW~Catalogue (\cite{MWC05})
}} \label{tab:new} {\tiny
\begin{supertabular}{lcccccccccc}
%C/1849 G2 & 1.159406 & 1.000940 & 2A  & {\bf --} &  --             &  --              & --              & --              \\
%%%C/1853 R1 & 0.172863 & 1.000664 &     & {\bf --} &  --             &  --              & --              & --              \\
C/1895 W1 & 0.192  & 18951118--18960810 &  114     & 1B  &  I  & $-146.0\pm$13.5 & $+201.1\pm$280.4 & $+482.4\pm$13.5 & $+430.1\pm$34.0  & pp   \\
C/1975 V2 & 0.219  & 19751113--19760209 &   85     & 2A  &  -- & $-138.8\pm$25.5 & $+358.0\pm$153.6 & $+1135\pm$26     & $+880.3\pm$122.9& PIdif\\
C/1911 S3 & 0.303  & 19110929--19120217 &  287     & 2A  &  -- & $-160.7\pm$42.9 & $-457.5\pm$260.3 & $+88.4\pm$42.9  & $+578.4\pm$315.2 & pp   \\
C/1899 E1 & 0.327  & 18990305--18990811 &  305     & 1B  &  I  & $-54.9\pm$12.5  & $-189.2\pm$61.3  & $-1199\pm$13    & $-1078\pm$38     & PIgr \\
C/1940 R2 & 0.368  & 19400825--19410401 &  370     & 1B  &  I  & $-29.6\pm$26.3  & $+44.0\pm$33.8   & $-1600\pm$26    & $-1386\pm$59     & pp   \\
C/1991 Y1 & 0.644  & 19911224--19920502 &  274     & 1B  &  I  & $-97.3\pm$8.8  & $-60.2\pm$10.3    & $+1111\pm$9      & $+1148\pm$10    & PIdif\\
C/2002 O4 & 0.776  & 20020727--20021001 & 1223     & 2A  &  -- & $-835.2\pm$19.6& $-692.5\pm$62.4   & $-371.3\pm$19.6  & $+29.4\pm$49.1  & pp   \\
C/1952 W1 & 0.778  & 19521210--19530718 &   36     & 1B  &  I  & $-140.9\pm$22.2 & $-1.1\pm$84.7    & $-298.8\pm$22.2 & $-42.3\pm$108.9  & PIgr \\
C/1975 X1 & 0.864  & 19751209--19760204 &   82     & 2B  &  -- & $-1053\pm$243    & $+805.1\pm$573.4 & $-1781\pm$243    & $77.3\pm$573.5 & PIdif\\
C/1955 O1 & 0.885  & 19550802--19551112 &   65     & 2A  &  -- & $-584.2\pm$15.2 & $+1892\pm$380    & $-289.0\pm$15.2 & $+1839\pm$348    & PImore\\
C/1996 N1 & 0.926  & 19960704--19961103 &  316     & 1B  &  AA & $-156.7\pm$6.9 &    --             & $+530.1\pm$6.9   & --              & PIdif\\
C/1993 Q1 & 0.967  & 19930816--19940417 &  526     & 1A  &  I  & $-1.1\pm$3.4   & $-5.4\pm$8.6      & $-70.1\pm$3.4    & $-223.0\pm$38.6 & pp   \\
C/1892 Q1 & 0.976  & 18920901--18930713 &  228     & 1B  &  I  & $-58.6\pm$12.7  & $+31.7\pm$19.3   & $-570.0\pm$12.7 & $-478.6\pm$19.3  & PIgr \\
C/1896 V1 & 1.063  & 18961103--18970430 &   91     & 1B  &  AA & $+0.5\pm$33.0   &   --             & $-804.1\pm$33.0 & --               & pp   \\
%C/1904 Y1 & 1.882049 & 1.000678 & 2A  & {\bf --} & $+2.6\pm$118.2  &   --             & $+248.4\pm$118.3& --              \\
%{\bf C/1906 B1}& {\bf 1.296435} & {\bf 1.000099} & {\bf 1B}  & {\bf --} &  --             &  --              & --              & --              \\
C/1996J1B & 1.298  & 19960510--19981217 &  529     & 1A  &  I  & $-1.4\pm$2.0   & $-10.4\pm$15.7    & $+568.0\pm$2.0   & $+603.8\pm$2.4  & PIgr \\
C/1996 E1 & 1.359  & 19960315--19961012 &  249     & 1A  &  I  & $-43.3\pm$4.2  & $+28.1\pm$11.5    & $+355.0\pm$4.2   & $+343.0\pm$31.3 & PaperII\\
C/1932 M1 & 1.647  & 19320621--19330120 &  187     & 1B  &  AA & $+19.1\pm$24.7  &    --            & $-250.9\pm$24.7 & --               & PImore\\
C/1946 C1 & 1.724  & 19460129--19470809 &  498     & 1A  &  I  & $-13.1\pm$5.3   & $+92.1\pm$19.8   & $+373.0\pm$5.3  & $+335.6\pm$9.4   & PIdif \\
C/1978 R3 & 1.772  & 19780914--19790925 &   52     & 1B  &  AA & $+67.6\pm$32.9 &    --             & $+283.7\pm$32.9  & --              & pp   \\
C/1983 O2 & 2.255  & 19830804--19840605 &   39     & 1B  &  AA & $-16.1\pm$20.9  &    --            & $+403.8\pm$20.9  & --              & PaperI\\
C/1898 V1 & 2.285  & 18981115--18990604 &   71     & 1B  &  AA & $-14.1\pm$59.5  &   --             & $+676.7\pm$59.4 & --               & PImore\\
C/1946 U1 & 2.408  & 19461101--19481002 &  143     & 1A  &  I  & $+2.7\pm$6.7    & $+192.1\pm$33.0  & $+29.5\pm$6.7   & $+12.2\pm$10.5   & PIdif\\
C/1987 W3 & 3.333  & 19870924--19900429 &   34     & 1A  &  AA & $+24.4\pm$7.3  &    --             & $-361.6\pm$7.3   & --              & pp   \\
C/1997BA6 & 3.436  & 19970111--20040815 &  529     & 1A  &  I  & $-1.5\pm$0.5   & $+38.4\pm$1.7     & $+371.8\pm$0.5   & $+404.9\pm$1.7  & PIgr \\
C/2002 R3 & 3.870  & 20020804--20031210 & 1268     & 1B  &  AA & $+39.0\pm$0.6  &    --             & $+9.9\pm$0.6     & --              & pp   \\
C/1942 C2 & 4.113  & 19420212--19430311 &   48     & 1A  &  AA & $-26.6\pm$14.7  &    --            & $-275.8\pm$14.7 & --               & PIdif\\
C/2002 A3 & 5.151  & 20020113--20030624 &  291     & 1B  &  AA & $+25.2\pm$27.8 &    --             & $+6179\pm$28     & --              & pp   \\
C/1978 G2 & 6.283  & 19780412--19800123 &    7     & 1B  &  AA & $-22.5\pm$45.0  &    --            & $-99.3\pm$45.0   & --              & PaperI\\
%C/1940 S1 & 1.061768 & 1.001459 & 2B  & {\bf --} & $+6140\pm$484   &    --            & $+5137\pm$484   & --              \\
%C/1959 O1 & 1.250534 & 1.002679 & 2B  & {\bf --} &  --             &    --            & --              & --               \\
%C/1968 N1 & 1.160434 & 1.000665 & 2A  & {\bf --} & $-149.6\pm$116.6&    --            & $+192.1\pm$116.8& --               \\
%C/1980 R1 & 2.112441 & 1.001723 & 2A  & {\bf --} & {\bf --} & $-12.9\pm$139.6 &    --            & $-361.7\pm$139.3 & --              \\
%C/1987 A1 & 0.921485 & 1.000383 & 2A  & {\bf --} & $-49.5\pm$123.6 &    --            & $-92.6\pm$123.4  & --              \\
%C/1996J1A & 1.297766 & 1.001572 & 1A  & \multicolumn{5}{c}{secondary component of the splitting}\\
%C/1997 P2 & 4.263334 & 1.028407 & 2B  & {\bf --} & {\bf --} & $-42.3\pm$18.0 &    --             & $-2368\pm$39     & --              \\
%%%C/2002 O4 & 0.776203 & 1.000855 & 2A$^{*}$          & {\bf --} & $-835.2\pm$19.6& $-692.5\pm$62.4   & $-371.3\pm$19.6  & $+29.4\pm$49.1  \\
%%%C/2002 R3 & 3.869597 & 1.002874 & 1B$^{*}$          & $+39.0\pm$0.6  &    --             & $+9.9\pm$0.6     & --              \\
%%%C/2002 T7 & 0.615750 & 1.000733 & NG orbit$^{*}$    & $-24.7\pm$0.3  & $+16.6\pm$0.4     & $-664.4\pm$0.3   & $-653.3\pm$1.1  \\
%%%C/2002 A3 & 5.151436 & 1.007890 & 1B$^{*}$          & $+25.2\pm$27.8 &    --             & $+6179\pm$28     & --              \\
%%%C/1986P1B & 1.199345 & 1.000797 & --  & \multicolumn{4}{c}{secondary component of the splitting} \\
&&&&&&&&&& \\
 \multicolumn{11}{c}{\bf  Oort spike comet for the Catalogue orbits; hyperbolic in the present paper}   \\
C/1991 X2 & 0.199  & 19911213--19920303 &  155     & 2B  &  -- & $-83.9\pm$174.9& $-26.3\pm$151.2& $-710.8\pm$174.9  & $-653.2\pm$151.1  & pp   \\
% \hline
&&&&&&&&&& \\
 \multicolumn{11}{c}{\bf (ii) ~~Long period comets with NG-effects in the MW~Catalogue}                       \\
 \multicolumn{11}{c}{\bf and $1/a_{\rm ori} < 10^{-5}$AU$^{-1}$ for orbit determined in the present paper}\\
C/1956 R1 & 0.316  & 19561108--19580411 &  249     & \multicolumn{1}{c}{NG orbit} &  I  & $-104.8\pm$3.1  & $+7.6\pm$11.5   & $-610.3\pm$5.1  & $-587.4\pm$10.9 & PImore \\
C/1959 Y1 & 0.504  & 19600104--19600617 &   88     & \multicolumn{1}{c}{NG orbit} &  I  & $-138.9\pm$16.4 & $+3.5\pm$143.4  & $-590.6\pm$16.4 & $-280.5\pm$49.4 & PaperII\\
C/2002 T7 & 0.616  & 20021012--20050512 & 4408     & \multicolumn{1}{c}{NG orbit} &  I  & $-24.7\pm$0.3   & $+16.6\pm$0.4   & $-664.4\pm$0.3  & $-653.3\pm$1.1  & pp     \\
C/1989 Q1 & 0.642  & 19890824--19891224 &  231     & \multicolumn{1}{c}{NG orbit} &  I  & $-4.3\pm$12.7   & $+62.5\pm$28.3  & $+190.4\pm$12.7 & $+92.8 \pm$38.3 & PIdif  \\
C/1885 X1 & 0.642  & 18851201--18860719 &  228     & \multicolumn{1}{c}{NG orbit} &  I  & $+8.7\pm$13.7   & $+31.3\pm$20.0  & $-245.3\pm$13.7 & $-100.4\pm$48.2 & pp     \\
C/2001 Q4 & 0.961  & 20010824--20050826 & 2552     & \multicolumn{1}{c}{NG orbit} &  I  & $+5.6\pm$0.5    & $+57.4\pm$0.8   & $-738.6\pm$0.5  & $-695.3\pm$0.8  & pp     \\
C/1995 Y1 & 1.055  & 19951226--19960921 &  262     & \multicolumn{1}{c}{NG orbit} &  I  & $-19.6\pm$6.3   & $-46.4\pm$14.9  & $+532.1\pm$6.3  & $+645.1\pm$10.7 & PaperII\\
C/1998 P1 & 1.147  & 19980811--19990515 &  461     & \multicolumn{1}{c}{NG orbit} &  I  & $-125.1\pm$10.1 & $+211.1\pm$11.1 & $+1119\pm$10    & $+1320\pm$8     & PaperII\\
C/1986 P1A& 1.200  & 19860805--19890411 &  688     & \multicolumn{1}{c}{NG orbit} &  I  & $-0.4\pm$1.0    & $+42.4\pm$2.4   & $+725.2\pm$1.0  & $+767.6\pm$2.8  & PaperII\\
C/1971 E1 & 1.233  & 19710309--19710909 &  138     & \multicolumn{1}{c}{NG orbit} &  I  & $-182.0\pm$72.4 & $+296.0\pm$128.9& $-527.8\pm$72.4 & $-97.8 \pm$71.6 & PIdif  \\
C/1993 A1 & 1.938  & 19930102--19940610 &  746     & \multicolumn{1}{c}{NG orbit} &  I  & $-19.3\pm$3.9   & $+57.9\pm$2.7   & $-539.3\pm$3.9  & $-408.6\pm$5.3  & PaperII\\
% \hline
&&&&&&&&&& \\
 \multicolumn{11}{c}{\bf (iii) ~~Oort spike comets for the catalogue orbits} \\
C/1989 X1 & 0.350  & 19891206--19900627 &  281     & 1A &  II & $+22.4\pm$2.4 & $+42.3\pm$12.8 & $-365.6\pm$2.4 & $-373.3\pm$6.9       & pp   \\
C/1990 K1 & 0.939  & 19900521--19920401 &  678     & NG orbit &  II & $+26.0\pm$1.6 & $+111.6\pm$7.2 & $-848.6\pm$1.6 & $-790.4\pm$3.4 & PaperII\\
C/1915 C1 & 1.005  & 19150213--19160926 &  646     & NG orbit &  II & $+73.1\pm$5.2 & $+224.9\pm$18.8& $+958.7\pm$5.2 & $+1042\pm$9    & pp   \\
C/2003 K4 & 1.024  & 20030528--20050903 & 3190     & 1A &  II & $+38.1\pm$0.2 & $+28.2\pm$0.6  & $-180.2\pm$0.2 & $-118.3\pm$1.2       & pp   \\
C/1913 Y1 & 1.106  & 19131218--19150907 & 1046     & 1A &  II & $+23.2\pm$1.7 & $+58.1\pm$5.7  & $+57.5\pm$1.7  & $+92.3\pm$5.7        & pp   \\
C/1978 H1 & 1.136  & 19780428--19791209 &  287     & 1A &  II & $+24.4\pm$2.7 & $+77.5\pm$17.3 & $-1028\pm$3    & $-1020\pm$8          & pp   \\
C/1947 Y1 & 1.500  & 19480110--19481130 &  124     & 1B &  II & $+27.3\pm$1.1 & $+181.1\pm$49.6& $+52.7\pm$1.1  & $+30.8\pm$12.9       & pp   \\
C/1997 J2 & 3.051  & 19970505--19991009 & 1446     & 1B &  II & $+36.8\pm$0.6 & $+45.5\pm$1.0  & $-4.8\pm$0.6   & $+15.5\pm$1.1        & PIgr \\
%C/2003 K4 & 1.024072 & 1.000427 & 1A$^{*}$ & {\bf II} & $+38.1\pm$0.2 & $+28.2\pm$0.6  & $-180.2\pm$0.2 & $-118.3\pm$1.2 \\
% \hline
&&&&&&&&&& \\
 \multicolumn{11}{c}{\bf (iv) ~~'Old' comets \bf with $1/a_{\rm ori} > 10^{-4}$AU$^{-1}$ for orbits determined in the present paper}   \\
C/1974 C1 & 0.503  & 19740214--19741118 &  198     & \multicolumn{1}{c}{NG orbit} &  III & $+697.2\pm$12.4 & $+563.9\pm$68.8 & $+710.3\pm$12.5 & $+785.8\pm$16.1 & pp   \\
C/2000 WM1& 0.555  & 20001116--20021013 & 2206     & \multicolumn{1}{c}{NG orbit} &  III & $+518.3\pm$0.4  & $+535.4\pm$1.2  & $-247.6\pm$0.4  & $-227.6\pm$1.3  & pp   \\
C/1985 R1 & 0.695  & 19850913--19860319 &  173     & \multicolumn{1}{c}{NG orbit} &  III & $+564.5\pm$3.6  & $+595.2\pm$20.9 & $+231.7\pm$5.6  & $+288.1\pm$17.9 & pp   \\
C/2003 T4 & 0.850  & 20031013--20050512 &  908     & \multicolumn{1}{c}{NG orbit} &  III & $+231.5\pm$3.5  & $+321.6\pm$1.9  & $-764.4\pm$3.5  & $-169.0\pm$1.7  & pp   \\
C/1953 T1 & 0.970  & 19531116--19540628 &   59     & \multicolumn{1}{c}{NG orbit} &  III & $+204.6\pm$17.7 & $+147.8\pm$23.3 & $-830.5\pm$17.4 & $-688.6\pm$23.3 & pp   \\
C/2002 U2 & 1.209  & 20021025--20030522 &  416     & 1B                           &  III & $+1033\pm$3     & $+1214\pm$53    & $+676.0\pm$3.0  & $+715.0\pm$5.3  & pp   \\
C/1998 M5 & 1.742  & 19980630--20000508 & 1222     & 1A                           &  III & $+512.8\pm$0.5  & $+510.9\pm$1.0  & $+2612.5\pm$0.4 & $+2628\pm$2     & pp   \\
% \hline
\end{supertabular}
}
%%%%\end{center}
%\end{table*}

\end{changemargin}

\end{document}